\documentclass[10pt,cite,epsf,epsfig]{article}
\usepackage{epsfig}
\usepackage{float}
\usepackage{cite}
\usepackage{amsmath}
\usepackage{amssymb}
\usepackage{epsfig,subfigure}
\usepackage{graphicx}
\usepackage{bbm}
\usepackage{enumerate}
\usepackage{wasysym}

\begin{document}
\author{S. Dev \thanks{sdev@associates.iucaa.in} $^{,1}$,
Desh Raj \thanks{raj.physics88@gmail.com} $^{,2}$, Radha Raman Gautam \thanks{gautamrrg@gmail.com} $^{,2}$, Lal Singh \thanks{lalsingh96@yahoo.com} $^{,2}$}
\date{$^1$\textit{Department of Physics, School of Sciences, HNBG Central University, Srinagar, Uttarakhand 246174, INDIA.}\\
\smallskip
$^2$\textit{Department of Physics, Himachal Pradesh University, Shimla 171005, INDIA.}}
\title{New mixing schemes for (3+1) neutrinos}
\maketitle
\begin{abstract}
We propose new mixing schemes for (3+1) neutrinos which describe mixing among active-active and active-sterile neutrinos. The mixing matrix in these mixing schemes can be factored into a zeroth order flavor symmetric part and another part representing small perturbations needed for generating non-zero $U_{e3}$, nonmaximal $\theta_{23}$, CP violation and active-sterile mixing. We find interesting correlations amongst various neutrino mixing angles and, also, calculate the parameter space for various parameters.
\end{abstract}

\section{Introduction}
The discovery of massive neutrinos by Super-Kamiokande experiment \cite{skam} in the year 1998 has paved the way for new physics beyond the Standard Model (BSM) of particle physics. Various neutrino oscillation experiments in the last two decades have measured the three neutrino mixing angles and two (atmospheric $|\Delta m_{23}^2|$ and solar $\Delta m_{21}^2$) mass splittings rather precisely. However, several anomalies at Short Base Line (SBL) neutrino experiments indicate $eV$ scale mass splitting. These anomalies were first reported by LSND experiment \cite{lsnd} in their anti-neutrino flux measurements and, subsequently, confirmed by MiniBooNE experiment \cite{miniboone} in both the neutrino and antineutrino modes. The recent MiniBooNE data \cite{miniboone2} also support these anomalies. In addition, reactor experiments \cite{reactor} and gallium solar neutrino experiments \cite{gallium} strongly support these anomalies. A possible explanation of these anomalies would require, at least, one $eV$ scale mass eigenstate in the neutrino sector and the decay width of Z boson would require the fourth neutrino state to be sterile. The recent global analysis \cite{mona} of neutrino oscillations in the presence of eV-scale sterile neutrinos, supports the explanation of reactor anomaly in terms of sterile neutrino oscillations in 3+1 scenario but disfavour sterile neutrino explanation of LSND anomaly. Reactor neutrino data favour sterile neutrino oscillation with $\Delta m_{41}^{2} \approx$ 1.3eV$^2$ and $|U_{e4}|\approx 0.1$ at the $3\sigma$ confidence level (CL)\cite{gariazzo, mona}.\\
 The recent Planck data \cite{planck} limit the effective number of relativistic degrees of freedom to $N_{eff}=3.15\pm0.23$ (Planck TT+lowP+BAO) at $95\%$ CL and the sum of neutrino masses to be $\sum m_{\nu}\leq0.23$ eV at the same confidence level. This is consistent with the bound given by the standard model of cosmology: $N_{eff}=3.046$. Although the cosmological bounds and latest Planck data disfavour the existence of $eV$ scale sterile neutrinos, this apparent conflict can be resolved by considering the presence of new BSM physics . In this context it has been proposed \cite{tang} that suppressed effective mixing can lead to a decrease of $N_{eff}$ if sterile neutrinos have self-interactions and the presence of additional gauge interactions can suppress the production of sterile neutrinos via flavor oscillations \cite{tram,das}. Therefore, there is still possibility for the existence of eV scale sterile neutrinos. Also, from the theoretical standpoint, the sterile neutrinos could be the obvious candidates for right-handed neutrinos in the Standard Model of particle physics.\\
Super-Kamiokande has provided upper bounds on sterile neutrino parameters $|U_{\mu4}|^{2}<0.041$ and $|U_{\tau 4}|^{2}<0.18$ at $90\%$ CL \cite{skc}. The recent data from reactor and other short and long baseline neutrino experiments such as MINOS\cite{minos}, Daya Bay\cite{daya} etc. provide new bounds on active-sterile mixing and $\Delta m_{41}^2$. Several ongoing and future long baseline experiments such as DUNE\cite{dune}, T2HK\cite{t2hk}, T2HKK\cite{t2hkk} etc. may shed more light on neutrino oscillation physics and explore active-sterile mixing. The phenomenology and experimental constraints on (3+1) neutrinos have been reviewed in \cite{kang,kyu,daijiro,ivan,leonard,diaz,kopp,li,parke,giunti,liu,gupta,tarak,rode,maria,zang,common,hong,diana,deba,model}.\\
The obvious next step is to build models that can give predictions on some neutrino parameters including the active-sterile non-trivial mixing and mass splitting of the order of $eV$ scale as indicated by recent neutrino oscillation experiments. A four neutrino (3+1) scheme for explaining neutrino masses and mixing has large number of free parameters. Models with flavor symmetries can lead to specific structures of neutrino mass and mixing matrix with reduced number of free parameters. Neutrino models based on flavor symmetries have been extensively employed to explain the mixing matrix within the three neutrino framework. In lepton mass models, the residual flavor symmetries may remain intact even if the original flavor symmetry of the Lagrangian is broken. These different residual symmetries lead to different mixing matrices for the charged lepton and neutrino sectors. Neutrino mixing matrices based on residual flavor symmetries viz. tribimaximal mixing (TBM) \cite{tbm}, bimaximal mixing (BM) \cite{bm}, hexagonal mixing (HM) \cite{hm}, democratic mixing (DM) \cite{dm}, golden ratio mixing I (GRM1) \cite{grm1} and golden ratio mixing II (GRM2) \cite{grm2} predict a vanishing reactor mixing angle ($\theta_{13}$) and a maximal atmospheric mixing angle ($\theta_{23}$) and, hence, need  modifications to satisfy the data from current neutrino oscillation experiments. Two mixing schemes TFH1 (Toorop-Feruglio-Hagedorn 1) and TFH2, proposed in Ref. \cite{tfh}, predict nonzero $\theta_{13}$ and nonmaximal $\theta_{23}$ and need corrections to explain the three mixing angles, simultaneously. 
The TBM mixing matrix is given by
\begin{equation}
U_{\text{TBM}}=
\left(
\begin{array}{ccc}
 \sqrt{\frac{2}{3}} & \frac{1}{\sqrt{3}} & 0 \\
 -\frac{1}{\sqrt{6}} & \frac{1}{\sqrt{3}} &
   -\frac{1}{\sqrt{2}} \\
 -\frac{1}{\sqrt{6}} & \frac{1}{\sqrt{3}} &
   \frac{1}{\sqrt{2}} \\
\end{array}
\right).
\end{equation}
Recent neutrino oscillation experiments have measured a non-zero $\theta_{13}$ which implies that (1,3) element of the neutrino mixing matrix ($U_{e3}$) cannot be zero. Thus we need to modify the TBM mixing matrix to accommodate a non-zero $\theta_{13}$. One simple possibility is to keep
one of the columns of the TBM mixing matrix intact while modifying its remaining two columns within the unitarity constraints.
This gives rise to three mixing patterns viz. $(U_{C1})_{TBM}$,
$(U_{C2})_{TBM}$ and $(U_{C3})_{TBM}$ which have their first, second and third columns identical to the TBM mixing matrix, respectively.
$(U_{C1})_{TBM}$ mixing is given by
\begin{equation}\label{eq:tm1}
(U_{C1})_{TBM}=\left(
\begin{array}{ccc}
  \frac{2}{\sqrt{3}} & \frac{1}{\sqrt{3}} \cos \theta &
\frac{1}{\sqrt{3}} \sin \theta \\
  -\frac{1}{\sqrt{6}} &
      \frac{1}{\sqrt{3}}\cos\theta+\frac{e^{i \phi} \sin
\theta}{\sqrt{2}} &
\frac{1}{\sqrt{3}}\sin\theta-
\frac{e^{i \phi} \cos\theta}{\sqrt{2}} \\
  -\frac{1}{\sqrt{6}} &
 \frac{1}{\sqrt{3}}\cos\theta-\frac{e^{i \phi}
                        \sin \theta}{\sqrt{2}} &
 \frac{1}{\sqrt{3}}\sin \theta
+\frac{e^{i \phi}
   \cos \theta}{\sqrt{2}}\end{array}
\right).
\end{equation}
$(U_{C1})_{TBM}$ reduces to the TBM mixing matrix in the special case $\theta=0$ and $\phi=0$. 
$(U_{C2})_{TBM}$ mixing is given by
\begin{equation}\label{eq:tm2}
(U_{C2})_{TBM}=\left(
\begin{array}{ccc}
 \sqrt{\frac{2}{3}} \cos \theta &
   \frac{1}{\sqrt{3}} & \sqrt{\frac{2}{3}}
   \sin \theta \\
 -\frac{\cos\theta}{\sqrt{6}}+\frac{e^{i \phi} \sin
\theta}{\sqrt{2}} & \frac{1}{\sqrt{3}} &
   -\frac{\sin\theta}{\sqrt{6}}-\frac{e^{i \phi} \cos\theta}{\sqrt{2}} \\
 -\frac{\cos\theta}{\sqrt{6}}-\frac{e^{i \phi}
   \sin \theta}{\sqrt{2}} &
   \frac{1}{\sqrt{3}} & -\frac{\sin
   \theta}{\sqrt{6}}
+\frac{e^{i \phi}
   \cos \theta}{\sqrt{2}}\end{array}
\right).
\end{equation}
$(U_{C2})_{TBM}$ mixing reduces to the TBM scheme in the special case 
$\theta=0$ and $\phi=0$. This mixing scheme is generally known as the trimaximal mixing.
$(U_{C3})_{TBM}$ mixing is given by
\begin{equation}\label{eq:tm3}
(U_{C3})_{TBM}=
\left(
\begin{array}{cccc}
\cos\theta & \sin\theta & 0 \\
-\frac{e^{-i\phi} \sin\theta}{\sqrt{2}} & \frac{e^{-i
  \phi} \cos\theta}{\sqrt{2}} & -\frac{1}{\sqrt{2}}  \\
-\frac{e^{-i\phi} \sin\theta}{\sqrt{2}} & \frac{e^{-i
  \phi} \cos\theta}{\sqrt{2}} & \frac{1}{\sqrt{2}}  \\
\end{array}
\right).
\end{equation}
$(U_{C3})_{TBM}$ reduces to the TBM mixing matrix in the special case 
$\theta=\arctan (1/\sqrt{2})$ and $\phi=0$.
The above partial mixing schemes have been successfully employed to explain the pattern of lepton mixing and have been extensively studied in the literature \cite{column, sanjeev}. Especially, the $(U_{C1})_{TBM}$ mixing gives a very good fit to the present neutrino oscillation data.\\
In  the present work, we present new mixing schemes for (3+1) neutrinos which are essentially partial mixing schemes having either one row or one column of the $4\times4$ mixing matrix to be the same as that of the popular mixing schemes like TBM, BM, DM, HM, GRM1, GRM2, TFH1 and TFH2. These mixing schemes accommodate active-active and active-sterile neutrino mixings. We, also, discuss general $4\times4$ mixing schemes with one column or one row fixed with none of the mixing matrix elements equal to zero. A $4\times4$ real mixing scheme with first or second column of the mixing matrix remaining the same as that of TBM has, already, been studied in Ref. \cite{dev} and it has been found that the mixing matrix with second column identical to TBM mixing matrix is the only viable case.\\
The plan of this paper is as follows. Sec. II describes the general (3+1) neutrino framework. In Sec. III, we present partial neutrino mixing schemes and study their phenomenology. Sec. IV describes the $(4\times4)$ general mixing schemes with one column/row fixed. Sec. V summarizes the main results of this work.
\section{The (3+1) neutrino framework}
The presence of sterile neutrino(s) affects the active neutrino mixing angles via the unitarity conditions of the mixing matrix i.e., $\Sigma_{j}\vert U_{ij}\vert^{2}=1$, where $i=e,\mu,\tau,s$ and $j=1,2,3,4$. Table 1 presents the bounds on active-sterile mixing matrix elements. The experimental mass splitting are $\Delta m^2_{21}=(7.05-8.14)\times 10^{-5}$ eV$^2$, $\Delta m^2_{31}=(2.43-2.67)\times 10^{-3}$ eV$^2$ for normal mass ordering and $(2.37-2.61)\times 10^{-3}$ eV$^2$ for inverted mass ordering at $3\sigma$ CL \cite{salas}. The fourth mass splitting is $\Delta m^2_{41}\approx1.7$ eV$^2$ (best fit)\cite{li}. The 3$\sigma$ ranges of the elements of the $3\times3$ sub matrix of $U$ with the bounds presented in \cite{parke} without imposing the unitarity of $U^{3\times3}$ are given by
\begin{eqnarray}
|U|^{3\times3}\equiv \left(
\begin{array}{ccc}
 0.76- 0.85 & 0.50- 0.60 & 0.13- 0.16 \\
 0.21- 0.54 & 0.42- 0.70 & 0.61- 0.79 \\
 0.18- 0.58 & 0.38- 0.72 & 0.40- 0.78
\end{array}
\right).
\end{eqnarray}
\begin{table}[h]
\begin{center}
 \begin{tabular}{ccc}
   \hline
   \hline
   Parameter & bound at $3 \sigma$ CL \\
   \hline
   $|U_{e4}|^2$  &  0.0098 $-$ 0.031\\
   $|U_{\mu4}|^2$  &  0.0060 $-$ 0.026\\
   $|U_{\tau4}|^2$  &  $\leq 0.039$ \\
    \hline
 \end{tabular}
 \end{center}
 \caption{The current experimental bounds on sterile neutrino mixing parameters Ref.\cite{li}}
\end{table}
In the four (3+1) neutrino framework, there are three active and one sterile neutrinos. The corresponding neutrino mixing matrix is a $4\times4$ unitary matrix. We use the following parametrization \cite{rode} for the $4\times4$ neutrino mixing matrix:
\begin{small}
\begin{equation}
U_{4\times 4}= R(\theta_{34}) R(\theta_{24},\delta_{24}) R(\theta_{14},\delta_{14}) R(\theta_{23}) R(\theta_{13},\delta_{13}) R(\theta_{12}) P
\end{equation}
\end{small}
where $R(\theta_{ij})$ matrix describes rotation in $ij^{\textrm{th}}$ plane and diagonal phase matrix $P$ contains three Majorana type CP-violating phases. The advantage of such parametrization is that for vanishing active-sterile mixing the above parametrization reduces to the standard PMNS parametrization for three active neutrinos. The six neutrino mixing angles in terms of mixing matrix elements can be written as 
\begin{small}
\begin{eqnarray}
\sin^2\theta_{14}&=& |U_{e4}|^2\nonumber,\\
\sin^2\theta_{24}&=& \frac{|U_{\mu4}|^2}{1-|U_{e4}^2|}\nonumber,\\
\sin^2\theta_{34}&=& \frac{|U_{\tau4}|^2}{1-|U_{e4}|^2-|U_{\mu4}|^2},\\
\sin^2\theta_{13}&=&\frac{|U_{e3}|^2}{1-|U_{e4}|^2}\nonumber,\\
\sin^2\theta_{12}&=& \frac{|U_{e3}|^2}{1-|U_{e4}|^2-|U_{e3}|^2}\nonumber,\\
\sin^2\theta_{23}&=& \frac{|U_{\mu 3}|^2 (1-|U_{e4}|^2)-|U_{e4}|^2 |U_{\mu 4}|^2}{(1-|U_{e4}|^{2}-|U_{\mu 4}|^{2})}+\frac{|U_{e1} U_{\mu1}+U_{e2} U_{\mu2}|^2 (1-|U_{e4}|^2)}{(1-|U_{e4}|^{2}-|U_{e 3}|^{2})(1-|U_{e4}|^{2}-|U_{\mu4}|^{2})}.\nonumber
\end{eqnarray}
\end{small}
\section{Partial Mixing Schemes}
 A partial mixing matrix $U_{Ci}$ ($U_{Ri}$) is defined as a $4\times 4$ unitary matrix with the $i$th column (row) fixed to be $N( a~~b~~1~~0)^T (N( 1~~b~~a~~0))$, while keeping other three columns (rows) free within the unitarity constraints. The parameters $a$ and $b$ for different mixing schemes have been summarized in Table II . $N=1/\sqrt{1+a^2+b^2}$ is the normalization constant. One can obtain a particular partial mixing matrix $U_{Ci}$ or $U_{Ri}$ by selecting the respective values of $a$ and $b$ listed in Table II. For example, choosing $a=2$ and $b=1$ for $U_{C1}$ leads to a $4\times 4$ unitary matrix with its first column identical to TBM mixing matrix. Fig 1 shows the parameter space for $a$ and $b$ under the current neutrino oscillation data for $U_{C1}, U_{C2}$ and $U_{R3}$ mixing schemes.
\begin{table}[h]
\begin{center}
\begin{tabular}{|c|c|c|c|}
 \hline \hline
 Mixing& $U_{C1} $ & $U_{C2}$ \\
pattern  &  $a~~~~~~~~~~~~~~~~~~~~~b$  &  $a~~~~~~~~~~~~~~~~~~~~b$ \\
\hline \hline
 TBM & 2~~~~~~~~~~~~~~~~~~~~~~~~1  &  1~~~~~~~~~~~~~~~~~~~~~~1 \\
 BM & $\sqrt{2}$~~~~~~~~~~~~~~~~~~~~~~1  &  $\sqrt{2}$~~~~~~~~~~~~~~~~~~~~1 \\
 DM & $\sqrt{\frac{3}{2}}$~~~~~~~~~~~~~~~~~~~~~$\frac{1}{\sqrt{2}}$  & $\sqrt{\frac{3}{2}}$~~~~~~~~~~~~~~~~~~~$\frac{1}{\sqrt{2}}$\\
 HM & $\sqrt{6}$~~~~~~~~~~~~~~~~~~~~~~1  &  $\sqrt{\frac{2}{3}}$~~~~~~~~~~~~~~~~~~~~1 \\
 GRM1 & $\sqrt{3+\sqrt{5}}$~~~~~~~~~~~~~~1  &   $\sqrt{3-\sqrt{5}}$~~~~~~~~~~~~~1 \\
 GRM2 & $\sqrt{2+\frac{4}{\sqrt{5}}}$~~~~~~~~~~~~~~1  & $\sqrt{10-4\sqrt{5}}$~~~~~~~~~~1 \\
TFH1 & ~~~~$\frac{1}{2}(\sqrt{3}+1)$~~~~~~~$\frac{1}{2}(\sqrt{3}-1)$ & 1 ~~~~~~~~~~~~~~~~~~~~~ 1 \\
TFH2 & ~~~$2+\sqrt{3}$~~~~~~~~~~~~$1+\sqrt{3}$ & 1~~~~~~~~~~~~~~~~~~~~~~~1\\   
 \hline \hline
 Mixing &  $U_{R2}$ & $U_{R3}$ \\
pattern  &  $a~~~~~~~~~~~~~~~~~~~~b$ &  $a~~~~~~~~~~~~~~~~~~~~b$  \\
\hline \hline
 TBM &  $\sqrt{3}$~~~~~~~~~~~~~~~~~~$\sqrt{2}$ & $\sqrt{3}$~~~~~~~~~~~~~~~~~~$\sqrt{2}$ \\
 BM & $\sqrt{2}$~~~~~~~~~~~~~~~~~~~~1 & $\sqrt{2}$~~~~~~~~~~~~~~~~~~~~1 \\
 DM &  2~~~~~~~~~~~~~~~~~~~~~~1 & 1~~~~~~~~~~~~~~~~~~~~~~1 \\
 HM &  1~~~~~~~~~~~~~~~~~~~~$\sqrt{3}$ & 1~~~~~~~~~~~~~~~~~~~~$\sqrt{3}$\\
 GRM1 & $\sqrt{\frac{1}{2}(5+\sqrt{5})}$~~~~~~$\sqrt{\frac{1}{2}(3+\sqrt{5})}$ & $\sqrt{\frac{1}{2}(5+\sqrt{5})}$~~~~~~$\sqrt{\frac{1}{2}(3+\sqrt{5})}$ \\
 GRM2 & $\sqrt{2+\frac{4}{\sqrt{5}}}$~~~~~~~~~~~~~$\frac{1+\sqrt{5}}{\sqrt{10-2\sqrt{5}}}$ & $\sqrt{2+\frac{4}{\sqrt{5}}}$~~~~~~~~~~~~~$\frac{1+\sqrt{5}}{\sqrt{10-2\sqrt{5}}}$\\
TFH1 & $2+\sqrt{3}$~~~~~~~~~~~~~~$1+\sqrt{3}$ & 1 ~~~~~~~~~~~~~~~~~~~ 1 \\
TFH2 & 1~~~~~~~~~~~~~~~~~~~~1 & $2+\sqrt{3}$~~~~~~~~~~~~~~$1+\sqrt{3}$\\  
 \hline \hline
 \end{tabular}
 \end{center}
 \caption{The values of the parameters $a$ and $b$ for the partial mixing schemes of type $U_{C1}, U_{C2}, U_{R2}$ and $U_{R3}$.}
\end{table}
\begin{figure}[h]
\begin{center}
\epsfig{file=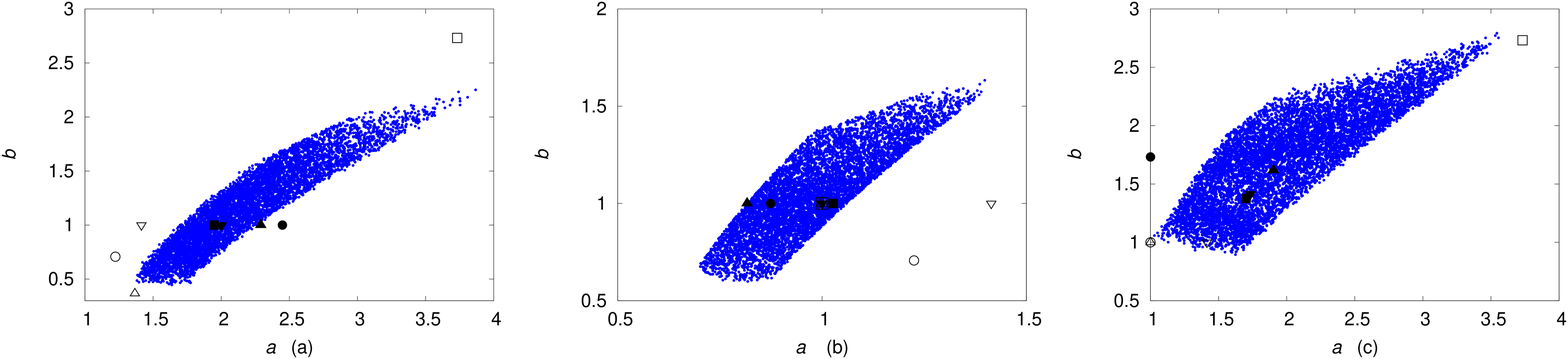, width=16.5cm, height=5.0cm}
\end{center}
\caption{The allowed parameter space for parameters $a$ and $b$ corresponding to parametrizations $U_{C1}$ (a), $U_{C2}$ (b), $U_{R3}$ (c) along with the values of $a$ and $b$ for TBM ($\blacktriangledown$), BM ($\bigtriangledown$), DM ($\bigcirc$), HM ($\bullet$), GRM1($\blacktriangle$), GRM2 ($\blacksquare$), TFH1 ($\bigtriangleup$) and TFH2 ($\square$). In case of $U_{C2}$ ($U_{R3}$), the values of $a$ and $b$ coincide for TBM, TFH1 and TFH2 (TFH1 and DM).}
\end{figure} 
\subsection{Mixing Scheme with first column fixed to $N(a~~b~~ 1~~ 0)^T$}
Here, we study the mixing scheme with first column fixed according to the well known mixing schemes:
\begin{eqnarray}
U_{C1}:\left(
\begin{array}{c}
U_{e1}\\
U_{\mu1}\\
U_{\tau1}\\
U_{s1}\\
\end{array}
\right)=\left(
\begin{array}{c}
a N\\
b N\\
N\\
0\\
\end{array}
\right).
\end{eqnarray}
The most general $4\times4$ mixing matrix with the first column fixed to be $N( a~~b~~ 1~~ 0)^T$, can be written as
\begin{small}
\begin{eqnarray}
U_{C1}=\left(
\begin{array}{cccc}
 a N & \sqrt{b^2+1} c_2 c_3 N & \sqrt{b^2+1} c_2 N s_3 &
   \sqrt{b^2+1} N s_2 \\
 -b N & \frac{\left(a b c_2 c_3 N+u\right)}{\sqrt{b^2+1}} & \frac{
   \left(a b c_2 N s_3+v\right)}{\sqrt{b^2+1}} & \frac{\left(a b N s_2-c_2 e^{i (\phi_2-\phi_1)}  s_1\right)}{\sqrt{b^2+1}} \\
 -N & \frac{\left(a c_2 c_3 N-b u\right)}{\sqrt{b^2+1}} & \frac{a c_2 N s_3-b v}{\sqrt{b^2+1}} & \frac{\left(b c_2 e^{i
   (\phi_2-\phi_1)} s_1+a N s_2\right)}{\sqrt{b^2+1}}
   \\
 0 & e^{i (\phi_1+\phi_3)} s_1 s_3-c_1 c_3 e^{i
   \phi_2} s_2 & -c_3 e^{i (\phi_1+\phi_3)}
   s_1-c_1 e^{i \phi_2} s_2 s_3 & c_1 c_2
   e^{i \phi_2} \\
\end{array}
\right)P
\end{eqnarray}
\end{small}
where $u=c_3 e^{i (\phi_2- \phi_1)} s_1 s_2+c_1 e^{i \phi_3} s_3,v=e^{i (\phi_2-\phi_1)} s_1 s_2 s_3-c_1 e^{i \phi_3} c_3$, $s_{i}=\sin\theta_{i}$ and $c_{i}=\cos\theta_{i}$. The phase matrix $P = $diag$(1,e^{i\alpha},e^{i\beta},e^{i\gamma})$ contains three Majorana phases. The values of $a$ and $b$ for different popular mixing schemes are summarized in Table II. The above matrix has been derived in the Appendix. Fixing one row or column of the mixing matrix provides three independent constraints on the mixing angles and CP-violating phases. Comparing the magnitudes of the elements of $U_{C1}$ mixing matrix with the unitary matrix in Eq. (6) imposes the following conditions:
\begin{equation}
|U_{e1}|=a N, ~~|U_{\mu1}|=b N~~ \textrm{and}~~ |U_{s1}|=0.\nonumber
\end{equation}
 The first condition $|U_{e1}|=a N$ gives
\begin{equation}
\cos^2\theta_{12}=\frac{a^2 N^2}{\cos^2\theta_{13} \cos^2\theta_{14}}=\frac{a^2 N^2}{1-|U_{e4}^2|-|U_{e3}^2|}\geq a^2 N^2.
\end{equation}
For $U_{C1}$ mixing, it is clear from Eq. (10) that $\theta_{12}$ decreases with increase in $\theta_{13}$ and $\theta_{14}$. Fig 1(a) shows that $U_{C1}$ mixing is viable only for TBM and GRM2 partial mixings. Also, $U_{C1}$ mixing predicts $\sin^2\theta_{12}\leq\frac{1}{3}$ for TBM and $\sin^2\theta_{12}\leq\frac{5-\sqrt{5}}{8}$ for GRM2.\\  
From the second condition $|U_{\mu1}|=b N$, we have
\begin{eqnarray}
b^2 N^2&=&|\cos\theta_{12} (\cos\theta_{13} \sin\theta_{14} \sin\theta_{24} \cos(\delta_{14}-\delta_{24})+\cos\delta_{13} \sin\theta_{13}
\sin\theta_{23} \cos\theta_{24})\nonumber\\
&&+\sin\theta_{12} \cos\theta_{23} \cos\theta_{24}|^2\\
&&+|\cos\theta_{12} (\cos\theta_{13} \sin\theta_{14} \sin\theta_{24} \sin(\delta_{14}-\delta_{24})+\sin\delta_{13} \sin\theta_{13}
\sin\theta_{23} \cos\theta_{24})|^2 \nonumber
\end{eqnarray}
and from third condition $|U_{s1}|=0$, we have
\begin{equation}
\tan\theta_{12}=|\frac{e^{i(\delta_{13}+\delta_{24})} \cos\theta_{13} \sin\theta_{14}-e^{i\delta_{14}}\sin\theta_{13} \left(\sin\theta_{23} \tan\theta_{24}+e^{i\delta_{24}} \cos\theta_{23} \sec\theta_{24} \tan\theta_{34}\right)}{e^{i(\delta_{13}+\delta_{14})} (\cos\theta_{23}
\tan\theta_{24}-e^{i\delta_{24}} \sin\theta_{23} \sec\theta_{24} \tan\theta_{34})}|.
\end{equation}
It is clear that the six mixing angles are not independent and are related as above. Using Eqs. (7) and (9), we obtain the following relations;
\begin{small}
\begin{eqnarray}
\sin^2\theta_{13}&=&\left(1-\frac{a^2 N^2}{\cos^2\theta_{14}}\right)\sin^2\theta_{3}\leq 1-a^2 N^2 (1+\sin^2\theta_{14})\nonumber,\\
\sin^2\theta_{12}&=&\frac{\left(b^2+1\right) \cos^2\theta_2 \cos^2\theta_3}{a^2+\left(b^2+1\right) \cos^2\theta_2 \cos^2\theta_3}\equiv 1-a^2 N^2 \sec^2\theta_{13} \sec^2\theta_{14}\nonumber,\\
\sin^2\theta_{23}&=&\frac{|U_{\mu3}|^2+|\cos_{12} U_{\mu 1}+\sin\theta_{12}U_{\mu 2}|^2-\sin^2\theta_{14}\sin^2\theta_{24}}{\cos^2\theta_{24}},\\
\sin^2\theta_{14}&=& \left(b^2+1\right) N^2 \sin^2\theta_2\nonumber,\\
\sin^2\theta_{24}&=&\frac{a^2 b^2 N^2 \sin^2\theta_{2}-2 a b N\sin\theta_{1}\sin\theta_{2} \cos\theta_{2} \cos(\phi_{1}-\phi_{2})+\sin^2\theta_{1} \cos^2\theta_{2}}{-\left(b^2+1\right)^2 N^2 \sin^2\theta_{2}+b^2+1}\nonumber,\\
\sin^2\theta_{34}&=&\frac{a N \left(a N \sin^2\theta_{2}+b\sin\theta_{1} \sin(2\theta_{2}) \cos(\phi_{1}-\phi_{2})\right)+b^2\sin^2\theta_{1}\cos^2\theta_{2}}{a b N \sin\theta_{1} \sin(2\theta_{2}) \cos(\phi_{1}-\phi_{2})-(b^2+(1+b^2)N^2)\sin^2\theta_{2}+b^2-\sin^2\theta_{1} \cos^2\theta_{2}+1}~.\nonumber
\end{eqnarray}
\end{small}
The mixing matrix in Eq. (9) can be factorized as
\begin{equation}
U_{C1}=V(a,b) R_{34}(\theta_{1},\phi_{1}) R_{24}(\theta_{2},\phi_{2}) R_{23}(\theta_{3},\phi_{3})P 
\end{equation}
where
\begin{equation}
V(a,b)=\left(
\begin{array}{cccc}
 a N & \sqrt{b^2+1} N & 0 & 0 \\
 -b N & \frac{a b N}{\sqrt{b^2+1}} & -\frac{1}{\sqrt{b^2+1}} & 0 \\
 -N & \frac{a N}{\sqrt{b^2+1}} & \frac{b}{\sqrt{b^2+1}} & 0 \\
 0 & 0 & 0 & 1 \\
\end{array}
\right).
\end{equation}
The matrix $V(a,b)$ represents one of the following mixing schemes: TBM, BM, DM, HM, GRM1, GRM2, TFH1, TFH2.  $V(a,b)$ reproduces different mixing schemes for different values of $a,b$ listed in Table II (except for TFH1 and TFH2). $R_{ij}$ denote small rotations in the $(ij)$ plane and represent perturbations to different mixing schemes. As an example, the partial mixing scheme with its first column fixed to the TBM values is obtained using Eqs. (9) and (14):
\begin{equation}
[U_{C1}]_{TBM}=V(2,1) R_{34}(\theta_{1},\phi_{1}) R_{24}(\theta_{2},\phi_{2}) R_{23}(\theta_{3},\phi_{3})P 
\end{equation} where
\begin{small}
\begin{eqnarray}
V(2,1)=\left(
\begin{array}{cccc}
 \sqrt{\frac{2}{3}} & \frac{1}{\sqrt{3}} & 0 & 0\\
 -\frac{1}{\sqrt{6}} & \frac{1}{\sqrt{3}} &
   -\frac{1}{\sqrt{2}} & 0\\
 -\frac{1}{\sqrt{6}} & \frac{1}{\sqrt{3}} &
   \frac{1}{\sqrt{2}} &0 \\
   0 & 0 & 0 & 1\\
\end{array}
\right),
R_{34}(\theta_{1},\phi_{1})=\left(
\begin{array}{cccc}
 1 & 0 & 0 & 0 \\
 0 & 1 & 0 & 0 \\
 0 & 0 & \cos\theta_1 & e^{-i \phi_1} \sin\theta_1 \\
 0 & 0 & -e^{i \phi_1} \sin\theta_1 & \cos\theta_1 \\
\end{array}
\right),\nonumber \\
 R_{24}(\theta_{2},\phi_{2}) = \left(
\begin{array}{cccc}
 1 & 0 & 0 & 0 \\
 0 & \cos\theta_2 & 0 & e^{-i \phi_2} \sin\theta_2 \\
 0 & 0 & 1 & 0 \\
 0 & -e^{i \phi_2} \sin\theta_2 & 0 & \cos\theta_2 \\
\end{array}
\right), R_{23}(\theta_{3},\phi_{3})=\left(
\begin{array}{cccc}
 1 & 0 & 0 & 0 \\
 0 & \cos\theta_3 & e^{-i\phi_3} \sin\theta_3 & 0 \\
 0 & -e^{i \phi_3} \sin\theta_3 & \cos\theta_3 & 0 \\
 0 & 0 & 0 & 1 \\
\end{array}
\right).
\end{eqnarray}
\end{small}
\begin{tiny}
\begin{equation}
\Rightarrow [U_{C1}]_{TBM}=\left(
\begin{array}{cccc}
 \sqrt{\frac{2}{3}} & \frac{c_2 c_3}{\sqrt{3}} & \frac{c_2 s_3}{\sqrt{3}} & \frac{s_2}{\sqrt{3}} \\
 -\frac{1}{\sqrt{6}} & \frac{c_2 c_3}{\sqrt{3}}+\frac{e^{-i \phi_1}
   \left(c_3 e^{i \phi_2} s_1 s_2+c_1 e^{i (\phi_1+\phi_3)} s_3\right)}{\sqrt{2}} & -\frac{c_1 e^{i \phi_3} c_3}{\sqrt{2}}+\frac{c_2 s_3}{\sqrt{3}}+\frac{e^{-i (\phi_1-\phi_2)} s_1 s_2 s_3}{\sqrt{2}} &
   \frac{s_2}{\sqrt{3}}-\frac{c_2 e^{-i (\phi_1-\phi_2)} s_1}{\sqrt{2}} \\
 -\frac{1}{\sqrt{6}} & \frac{c_2 c_3}{\sqrt{3}}+\frac{e^{-i \phi_1}
   \left(-c_3 e^{i \phi_2} s_1 s_2-c_1 e^{i (\phi_1+\phi_3)} s_3\right)}{\sqrt{2}} & \frac{c_1 e^{i \phi_3} c_3}{\sqrt{2}}+\frac{c_2 s_3}{\sqrt{3}}-\frac{e^{-i (\phi_1-\phi_2)} s_1 s_2 s_3}{\sqrt{2}} & \frac{c_2 e^{-i
   (\phi_1-\phi_2)} s_1}{\sqrt{2}}+\frac{s_2}{\sqrt{3}} \\
 0 & e^{i (\phi_1+\phi_3)} s_1 s_3-c_1 c_3 e^{i\phi_2} s_2 & -c_3 e^{i (\phi_1+\phi_3)}
  s_1-c_1 e^{i \phi_2} s_2 s_3 & c_1 c_2   e^{i \phi_2} \\
\end{array}
\right)P.
\end{equation}
\end{tiny}

For numerical analysis, we generate about $10^{7}\sim10^{8}$ points randomly. We vary the parameters $\theta_{1},\theta_{2},\theta_{3}$ and $\phi_{1},\phi_{2},\phi_{3}$ within the ranges $(0-\frac{\pi}{2})$ and $(0-2\pi)$, respectively. Parameters $a$ and $b$ are chosen from Table II corresponding to the different partial mixing schemes. The experimental constraints on neutrino parameters from neutrino oscillation experiments are summarized in Eq. (5) and Table I which have been used to check the viability of above partial mixing schemes. Only TBM and GRM2 partial mixing schemes are allowed at 3$\sigma$ CL For $U_{C1}$. Fig. 2 shows the correlations among different neutrino mixing angles for $U_{C1}$ mixing scheme with first column fixed to be the same as that of TBM. The correlation plot shown in Fig. 2(a) between $\theta_{12}$ and $\theta_{13}$ is in the form of band (in contrast to a line in the three neutrino case) due to the presence of extra parameters from sterile sector. $\theta_{12}$ varies inversely with $\theta_{14}$ (Fig. 2(b)) which is also clear from Eq. (10).\\   
In the context of symmetry, the origin of the first eigenvector fixed as $N(a~~b~~1~~0)^T$, can be seen as the invariance of the neutrino mass matrix $M_{\nu}^{4\times4}$ under a $Z_{2}$ symmetry: $G_{1}^{T}M_{\nu}^{4\times4}G_{1} = M_{\nu}^{4\times4}$ where the $Z_2$ symmetry generator $G_{1}$ is defined as
\begin{eqnarray}
G_{1}&&=u_{1} u_{1}^{\dagger}-u_{2} u_{2}^{\dagger}-u_{3} u_{3}^{\dagger}-u_{4} u_{4}^{\dagger}\nonumber\\
&&=\left(
\begin{array}{cccc}
 \left(a^2-b^2-1\right) N^2 & -2 a b N^2 & -2 a N^2 & 0 \\
 -2 a b N^2 & \frac{b^2 \left(-a^2+b^2+1\right) N^2-1}{b^2+1} & \frac{b \left(\left(-a^2+b^2+1\right) N^2+1\right)}{b^2+1} & 0 \\
 -2 a N^2 & \frac{b \left(\left(-a^2+b^2+1\right) N^2+1\right)}{b^2+1} & \frac{\left(-a^2+b^2+1\right) N^2-b^2}{b^2+1} & 0 \\
 0 & 0 & 0 & -1 \\
\end{array}
\right).
\end{eqnarray} 
For the $U_{C1}$ partial mixing corresponding to TBM, the generator $G_{1}$ and the corresponding mass matrix are given by 
\begin{small}
\begin{equation}
G_{1}=\frac{1}{3}\left(
\begin{array}{cccc}
 1 & -2 & -2 & 0 \\
 -2 & -2 & 1 & 0 \\
 -2 & 1 & -2 & 0 \\
 0 & 0 & 0 & -3 \\
\end{array}
\right)~ 
\textrm{and}~M_{\nu}^{4\times4}=\left(
\begin{array}{cccc}
 x & y & z & \frac{f+g}{2} \\
 y & p+2 y-2 z & t & f \\
 z & t & p & g \\
 \frac{f+g}{2} & f & g & s \\
\end{array}
\right)
\end{equation}
\end{small}
where $t=-p+x-\frac{y}{2}+\frac{3 z}{2}$.
Similarly, for the $U_{C1}$ partial mixing corresponding to GRM2, we have
\begin{small}
\begin{eqnarray}
G_{1}&&=\left(
\begin{array}{cccc}
 \frac{1}{4} \left(-1+\sqrt{5}\right) & -\frac{1}{4} \sqrt{5+\sqrt{5}} & -\frac{1}{4} \sqrt{5+\sqrt{5}} & 0 \\
 -\frac{1}{4} \sqrt{5+\sqrt{5}} & \frac{1}{8} \left(-3-\sqrt{5}\right) & \frac{1}{8} \left(5-\sqrt{5}\right) & 0 \\
 -\frac{1}{4} \sqrt{5+\sqrt{5}} & \frac{1}{8} \left(5-\sqrt{5}\right) & \frac{1}{8} \left(-3-\sqrt{5}\right) & 0 \\
 0 & 0 & 0 & -1 \\
\end{array}
\right),\nonumber\\
\textrm{and}~M_{\nu}^{4\times4}&&=\left(
\begin{array}{cccc}
 t & y & z & \sqrt{\frac{5}{2}-\sqrt{5}}
   (f+g) \\
 y & x+\sqrt{2+\frac{4}{\sqrt{5}}} (y-z) & p & f \\
 z & p & x & g \\
 \sqrt{\frac{5}{2}-\sqrt{5}} (f+g) & f & g & s \\
\end{array}
\right),
\end{eqnarray}
\end{small}
where $t=p+x+\frac{1}{20} \sqrt{5+\sqrt{5}} \left(\left(5-7 \sqrt{5}\right) z-5 \left(-3+\sqrt{5}\right) y\right)$.
\begin{figure}[h]
\begin{center}
\epsfig{file=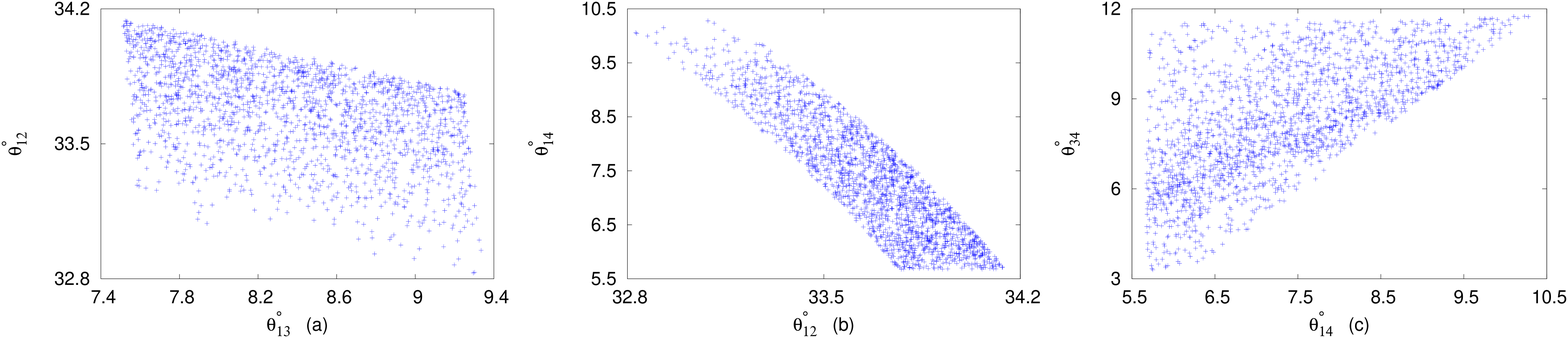, width=16.5cm, height=5.0cm}
\end{center}
\caption{Correlation plots among various neutrino oscillation parameters for parametrization $U_{C1}$ with TBM as partial flavour symmetry.}
\end{figure}
\subsection{Mixing Scheme with second column fixed to $N(a~~b~~ 1~~ 0)^T$}
The general mixing scheme with the second column fixed to $N( a~~b~~ 1~~ 0)^T$ is given by
\begin{small}
\begin{eqnarray}
U_{C2}=\left(
\begin{array}{cccc}
 \sqrt{b^2+1} c_2 c_3 N & a N & \sqrt{b^2+1} c_2 N s_3 & \sqrt{b^2+1} N s_2 \\
 \frac{\left(-a b c_2 c_3 N+u \right)}{\sqrt{b^2+1}} & b N & \frac{\left(-a b c_2 N s_3+v \right)}{\sqrt{b^2+1}} & -\frac{ \left(c_2 e^{i(\phi_2-\phi_1)} s_1+a b N s_2\right)}{\sqrt{b^2+1}} \\
 \frac{\left(-a c_2 c_3 N-b u\right)}{\sqrt{b^2+1}} & N & \frac{\left(-a c_2 N s_3-b v \right)}{\sqrt{b^2+1}} & \frac{\left(b c_2 e^{i (\phi_2-\phi_1)} s_1-a N s_2\right)}{\sqrt{b^2+1}} \\
 e^{i (\phi_1+\phi_3)} s_1 s_3-c_1 c_3 e^{i \phi_2} s_2 & 0 & -c_3 e^{i (\phi_1+\phi_3)}
s_1-c_1 e^{i \phi_2} s_2 s_3 & c_1 c_2 e^{i \phi_2} \\
\end{array}
\right)P
\end{eqnarray}
\end{small}
where $u=c_3 e^{i (\phi_2-\phi_1)} s_1 s_2+c_1 e^{i \phi_3} s_3, v=e^{i (\phi_2-\phi_1)} s_1 s_2 s_3-c_1 e^{i \phi_3} c_3$. From the condition $|U_{e2}|=a N$, one finds
\begin{equation}
\sin^2\theta_{12}=\frac{a^2 N^2}{\cos^2\theta_{13} \cos^2\theta_{14}}=\frac{a^2 N^2}{1-|U_{e4}^2|-|U_{e3}^2|}\geq a^2 N^2.
\end{equation}
$U_{C2}$ mixing for TBM, TFH1 and TFH2 partial mixings predicts $\sin^2\theta_{12}\geq\frac{1}{3}$. For HM, GRM1 and  GRM2 partial mixings $\sin^2\theta_{12}\geq\frac{1}{4}$, $\frac{5-\sqrt{5}}{10}$ and $\frac{5-\sqrt{5}}{8}$, respectively. Eq. (23) implies, $\theta_{12}$ increases with increase in $\theta_{13}$ and $\theta_{14}$ which is opposite to $U_{C1}$ mixing. The second condition $|U_{\mu2}|=b N$ implies
\begin{eqnarray}
b^2 N^2&=&|\cos\theta_{12} \cos\theta_{23} \cos\theta_{24}-\sin\theta_{12} (\cos\theta_{13}
   \sin\theta_{14} \sin\theta_{24} \cos(\delta_{14}-\delta_{24})\nonumber\\
   &&+\cos\delta_{13} \sin\theta_{13} \sin\theta_{23} \cos\theta_{24})|^2\nonumber\\
   &&+|\sin\theta_{12} (\cos\theta_{13} \sin\theta_{14} \sin\theta_{24} \sin(\delta_{14}-\delta_{24})+\sin\delta_{13} \sin\theta_{13}
\sin\theta_{23} \cos\theta_{24})|^2
\end{eqnarray}
and from third condition $|U_{s2}|=0$, we have
\begin{eqnarray}
\tan\theta_{12}=|\frac{e^{i(\delta_{13}+\delta_{14})} \left(\cos\theta_{23} \tan\theta_{24}-e^{i\delta_{24}} \sin\theta_{23} \sec\theta_{24} \tan\theta_{34}\right)}{e^{i\delta_{14}} \sin\theta_{13} \left(\sin\theta_{23}
\tan\theta_{24}+e^{i\delta_{24}} \cos\theta_{23} \sec\theta_{24} \tan\theta_{34}\right)-e^{i(\delta_{13}+\delta_{24})} \cos\theta_{13}\sin\theta_{14}}|.
\end{eqnarray}
Further from Eqs. (7) and (22), we have the following relations for mixing angles:
\begin{eqnarray}
\sin^2\theta_{13}&=&(1-\frac{a^2}{(1+a^2+b^2)\cos^2\theta_{14}})\sin^2\theta_{3}\leq 1-\frac{a^2 (1+\sin^2\theta_{14})}{1+a^2+b^2}\nonumber,\\
\sin^2\theta_{12}&=&\frac{a^2}{a^2+\left(b^2+1\right) \cos^2\theta_2 \cos^2\theta_3}\equiv \frac{a^2 \sec^2\theta_{13}\sec^2\theta_{14}}{1+a^2+b^2},\nonumber\\
\sin^2\theta_{14}&=& \left(b^2+1\right) N^2 \sin^2\theta_2\nonumber,\\
\sin^2\theta_{24}&=& \frac{a b N \left(a b N \sin^2\theta_2+\sin\theta_1 \sin(2\theta_2) \cos(\phi_1-\phi_2)\right)+\sin^2\theta_1 \cos^2\theta_2}{-\left(b^2+1\right)^2 N^2 \sin^2\theta_2+b^2+1}\nonumber,\\
\sin^2\theta_{34}&=&\frac{a N \left(b \sin\theta_1 \sin(2\theta_2)\cos(\phi_1-\phi_2)-a N \sin^2\theta_2\right)-b^2 \sin^2\theta_1) \cos^2\theta_2}{a b N \sin\theta_1 \sin(2\theta_2) \cos(\phi_1-\phi_2)+\left(b^2 \left(N^2+1\right)+N^2\right) \sin^2\theta_2-b^2+\sin^2\theta_1 \cos^2\theta_2-1}.\nonumber
\end{eqnarray}
It is clear that the neutrino mixing angle $\theta_{12}$ corresponding to mixing schemes $U_{C1}$ and $U_{C2}$ are related as  
\begin{equation}
\tan\theta_{12}|_{U_{C2}}=\frac{1}{\tan\theta_{12}|_{U_{C1}}}.
\end{equation} 
The $U_{C2}$ parametrization can be factorized as
\begin{equation}
U_{C2}=V(a,b) R_{34}(\theta_{1},\phi_{1}) R_{14}(\theta_{2},\phi_{2}) R_{13}(\theta_{3},\phi_{3}) P \nonumber
\end{equation}
where $R_{ij}$ denote complex  rotations in the $(ij)$ plane, $P=$ diag$\lbrace1,e^{i\alpha},e^{i\beta},e^{i\gamma}\rbrace$ is the phase matrix and $V(a,b)$ given by
\begin{equation}
V(a,b)=\left(
\begin{array}{cccc}
 \sqrt{b^2+1} N & a N & 0 & 0 \\
 -\frac{a b N}{\sqrt{b^2+1}} & b N & -\frac{1}{\sqrt{b^2+1}} & 0 \\
 -\frac{a N}{\sqrt{b^2+1}} & N & \frac{b}{\sqrt{b^2+1}} & 0 \\
 0 & 0 & 0 & 1 \\
\end{array}
\right)
\end{equation}
represents one of the mixing schemes such as TBM etc.\\
For numerical analysis, the parameters $\theta_{1},\theta_{2},\theta_{3}$ and $\phi_{1},\phi_{2},\phi_{3}$ are varied randomly within the ranges $(0 -\frac{\pi}{2})$ and $(0 - 2\pi)$, respectively. $a$ and $b$ are chosen from Table II corresponding to a particular partial mixing scheme and experimental constraints on neutrino oscillation parameters are used to check the viability of these partial mixing schemes. For $U_{C2}$ mixing schemes TBM, TFH1, TFH2, HM, GRM1 and GRM2 partial mixings are allowed at 3$\sigma$ CL. For $U_{C2}$ scheme there are almost similar correlations among neutrino oscillation parameters for all viable partial mixings and in Fig. 3, we have plotted correlations among neutrino mixing angles for TBM partial mixing.\\   
The generator $G_{2}$ corresponding to mass matrix $M_{\nu}^{4\times4}$ which leads to a mixing scheme with the second column fixed to $N( a~~b~~ 1~~ 0)^T$ is given by
\begin{small}
\begin{equation}
G_{2}=\left(
\begin{array}{cccc}
 \left(a^2-b^2-1\right) N^2 & 2 a b N^2 & 2 a N^2 & 0 \\
 2 a b N^2 & \frac{b^2 \left(-a^2+b^2+1\right) N^2-1}{b^2+1} & \frac{b \left(\left(-a^2+b^2+1\right) N^2+1\right)}{b^2+1} & 0 \\
 2 a N^2 & \frac{b \left(\left(-a^2+b^2+1\right) N^2+1\right)}{b^2+1} & \frac{\left(-a^2+b^2+1\right) N^2-b^2}{b^2+1} & 0 \\
 0 & 0 & 0 & -1 \\
\end{array}
\right).
\end{equation}
\end{small}
For TBM, TFH1 and TFH2 partial mixing schemes, we have
\begin{small}
\begin{eqnarray}
G_{2}=\frac{1}{3}\left(
\begin{array}{cccc}
 -1 & 2 & 2 & 0 \\
 2 & -1 & 2 & 0 \\
 2 & 2 & -1 & 0 \\
 0 & 0 & 0 & -3 \\
\end{array}
\right)~\textrm{and}~M_{\nu}^{4\times4}=\left(
\begin{array}{cccc}
 x & y & z & -f-g \\
 y & p-y+z & -p+x+y & f \\
 z & -p+x+y & p & g \\
 -f-g & f & g & s \\
\end{array}
\right). 
\end{eqnarray}
\end{small}
For HM partial mixing, we have
\begin{small}
\begin{eqnarray}
G_{2}&=&\frac{1}{4}\left(
\begin{array}{cccc}
 -2 & \sqrt{6} & \sqrt{6} & 0 \\
 \sqrt{6} & -1 & 3 & 0 \\
 \sqrt{6} & 3 & -1 & 0 \\
 0 & 0 & 0 & -4 \\
\end{array}
\right)\nonumber\\
\textrm{and}~
 M_{\nu}^{4\times4}&=&\left(
\begin{array}{cccc}
 x & y & z & -\sqrt{\frac{3}{2}} (f+g) \\
 y & p+\sqrt{\frac{2}{3}} (z-y) & -p+x+\frac{3 y+z}{\sqrt{6}} & f \\
 z & -p+x+\frac{3 y+z}{\sqrt{6}} & p & g \\
 -\sqrt{\frac{3}{2}} (f+g) & f & g & s \\
\end{array}
\right).
\end{eqnarray}
\end{small}
For GRM1 partial mixing
\begin{small}
\begin{eqnarray}
G_{2}=\frac{1}{-5 +\sqrt{5}}\left(
\begin{array}{cccc}
 -1+\sqrt{5} & \sqrt{2}-\sqrt{10} & \sqrt{2}-\sqrt{10} & 0 \\
 \sqrt{2}-\sqrt{10} & 3-\sqrt{5} & -2 & 0 \\
 \sqrt{2}-\sqrt{10} & -2 & 3-\sqrt{5} & 0 \\
 0 & 0 & 0 & 5-\sqrt{5} \\
\end{array}
\right) \nonumber\\
\textrm{and}~
 M_{\nu}^{4\times4}=\left(
\begin{array}{cccc}
 x & y & z & -\frac{f+g}{\sqrt{3-\sqrt{5}}} \\
 y & p+\sqrt{3-\sqrt{5}} (z-y) & \frac{t}{-130+58 \sqrt{5}} & f \\
 z & \frac{t}{-130+58 \sqrt{5}} & p & g \\
 -\frac{f+g}{\sqrt{3-\sqrt{5}}} & f & g & s \\
\end{array}
\right)
\end{eqnarray}
\end{small}
where $t=\left(130-58 \sqrt{5}\right) p+2 \left(29 \sqrt{5}-65\right) x+\sqrt{3-\sqrt{5}} \left(11 \sqrt{5} y-25 y-47 \sqrt{5} z+105 z\right)$.\\
For GRM2 partial mixing, the generator and the mass matrix are given by
\begin{small}
\begin{eqnarray}
G_{2}=\frac{1}{-3 + \sqrt{5}} \left(
\begin{array}{cccc}
 -2+\sqrt{5} & -\sqrt{\frac{5}{2}-\sqrt{5}} & -\sqrt{\frac{5}{2}-\sqrt{5}} & 0 \\
 -\sqrt{\frac{5}{2}-\sqrt{5}} & \frac{5}{2}-\sqrt{5} & -\frac{1}{2} & 0 \\
 -\sqrt{\frac{5}{2}-\sqrt{5}} & -\frac{1}{2} & \frac{5}{2}-\sqrt{5} & 0 \\
 0 & 0 & 0 & 3-\sqrt{5} \\
\end{array}
\right) \nonumber\\
 \textrm{and}~
 M_{\nu}^{4\times4}=\left(
\begin{array}{cccc}
 x & y & z & -\frac{f+g}{\sqrt{10-4 \sqrt{5}}} \\
 y & p+\sqrt{10-4 \sqrt{5}} (z-y) & \frac{t}{-170+76 \sqrt{5}} & f \\
 z & \frac{t}{-170+76 \sqrt{5}} & p & g \\
 -\frac{f+g}{\sqrt{10-4 \sqrt{5}}} & f & g & s \\
\end{array}
\right)
\end{eqnarray}
\end{small}
where $t=\left(170-76 \sqrt{5}\right) p+2 \left(38 \sqrt{5}-85\right) x+\sqrt{10-4 \sqrt{5}} \left(4 \sqrt{5} y-9 y-72 \sqrt{5} z+161 z\right)$.
\begin{figure}[h]
\begin{center}
\epsfig{file=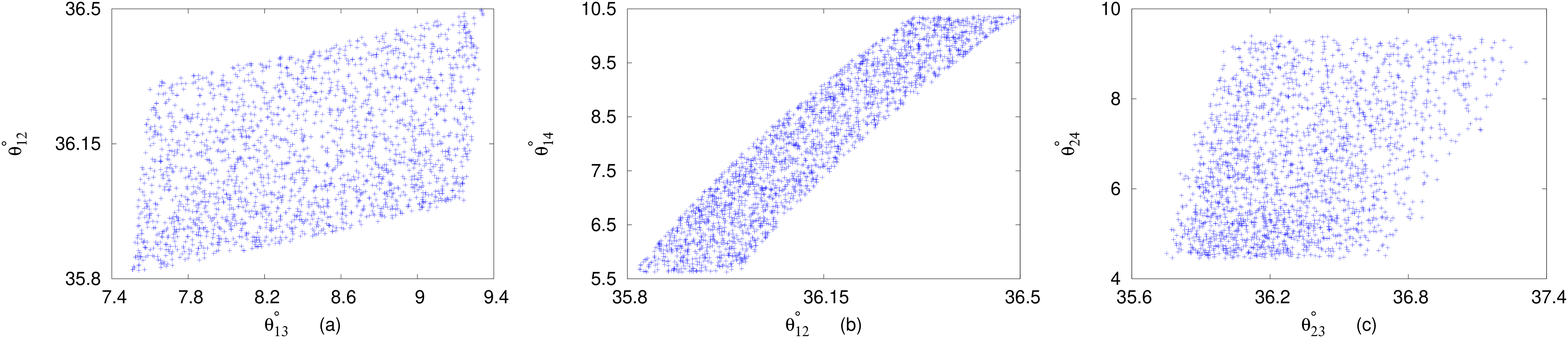, width=16.5cm, height=5.0cm}
\end{center}
\caption{Correlation plots among various neutrino oscillation parameters for $U_{C2}$ with TBM as partial flavour symmetry.}
\end{figure}
\subsection{Mixing Scheme with third row fixed to $N( 1~~b~~a~~ 0)$}
Here, we discuss the mixing scheme where the third row of the mixing matrix is fixed to $N( 1~~b~~a~~ 0)$. The first condition with $|U_{\tau 4}|=0$, implies
\begin{equation}
|\sin\theta_{34}|=0.
\end{equation}
The second condition $|U_{\tau 3}|=a N$ gives
\begin{equation}
\sin^2\theta_{23}=1-\frac{a^2 N^2}{\cos^2\theta_{13}}=1-\frac{a^2 N^2(1-|U_{e4}|^2)}{1-|U_{e4}|^2-|U_{e3}|^2}
\end{equation}
which limits $\sin^2\theta_{23}\leq 0.56$ for TBM and GRM1, $\sin^2\theta_{23}\leq \frac{1}{2}$ for BM and $\sin^2\theta_{23}\leq0.65$ for GRM2 partial mixings. From third condition $|U_{\tau 2}|=b N$, we have
\begin{equation}
\cos\delta_{13}=2 b^2 N^2 \csc(2\theta_{12}) \csc\theta_{13} \csc(2\theta_{23})-\frac{1}{2} \tan\theta_{12} \sin\theta_{13} \cot\theta_{t23}-\frac{1}{2} \cot\theta_{12} \csc\theta_{13} \tan\theta_{23}.
\end{equation}
The general mixing scheme with third row fixed to $N( 1~~b~~a~~ 0)$ can be parametrized as
\begin{small}
\begin{eqnarray}
U_{R3}=\left(
\begin{array}{cccc}
 \frac{c_2 e^{i \phi_2} \left(b c_1 e^{i \phi_1}-a N
   s_1\right)}{\sqrt{b^2+1}} & \frac{c_2 e^{i \phi_2} \left(e^{i
\phi_1} c_1+a b N s_1\right)}{\sqrt{b^2+1}} & -\sqrt{b^2+1}
   c_2 e^{i \phi_2} N s_1 & s_2 \\
 \frac{b u-a N x}{\sqrt{b^2+1}} &
   \frac{u+a b N x}{\sqrt{b^2+1}} &
   -\sqrt{b^2+1} N x & c_2 s_3 \\
 -N & b N & a N & 0 \\
 \frac{b v+a N y}{\sqrt{b^2+1}} &
   \frac{v-a b N y}{\sqrt{b^2+1}} &
   \sqrt{b^2+1} N y & c_2 c_3 \\
\end{array}
\right),
\end{eqnarray}
\end{small}
where $u=-c_3 e^{i (\phi_1+\phi_3)} s_1- c_1 e^{i
   (\phi_1+\phi_2)} s_2 s_3, v=-c_1 c_3 e^{i (\phi_1+\phi_2)} s_2+e^{i
   (\phi_1+\phi_3)} s_1 s_3, x=c_1 c_3 e^{i \phi_3}-s_1 s_2 s_3 e^{i \phi_2}$ and $y=c_1 s_3 e^{i \phi_3}+ c_3 s_1 s_2 e^{i \phi_2}$.
The $U_{R3}$ mixing scheme can be factorized as
\begin{equation}
U_{R3}=P^{\prime} R_{24}(\theta_{3},\phi_{3}) R_{14}(\theta_{2},\phi_{2}) R_{12}(\theta_{1},\phi_{1})V(a,b) P \nonumber
\end{equation}
where $R_{ij}$ are complex  rotations in the $(ij)$ plane while $P$ and $P^{\prime}$ are phase matrices and $V(a,b)$ given by
\begin{equation}
V(a,b)=\left(
\begin{array}{cccc}
 \frac{b}{\sqrt{b^2+1}} & \frac{1}{\sqrt{b^2+1}} & 0 & 0 \\
 -\frac{a N}{\sqrt{b^2+1}} & \frac{a b N}{\sqrt{b^2+1}} & -\sqrt{b^2+1} N & 0 \\
 -N & b N & a N & 0 \\
 0 & 0 & 0 & 1 \\
\end{array}
\right)
\end{equation}
reproduces different mixing schemes such as TBM for different values of $a$ and $b$.
For $U_{R3}$ mixing scheme we find following relations:
\begin{eqnarray}
\sin^2\theta_{14}&=& \sin^2\theta_2\nonumber,\\
\sin^2\theta_{24}&=& \sin^2\theta_{3}\nonumber,\\
\sin^2\theta_{34}&=& 0 \nonumber,\\
\sin^2\theta_{13}&=& N^2 (b^2+1)\sin^2\theta_{1}\nonumber,\\
\sin^2\theta_{12}&=&\frac{\cos^2\theta_{14}\left(a b N \left(a b N \sin^2\theta_{1}+\sin(2\theta_{1})
   \cos\phi_1\right)+\cos^2\theta_{1}\right)}{b^2+1}.
\end{eqnarray}
For the numerical analysis, we generate the parameters $\theta_{1},\theta_{2},\theta_{3}$ and $\phi_{1},\phi_{2},\phi_{3}$ randomly within the ranges $(0 - \frac{\pi}{2}$) and $(0 - 2\pi$), respectively. The parameters $a$ and $b$ are chosen from Table II and available experimental constraints on neutrino oscillation parameters are imposed to check the phenomenological viability of these mixing schemes. Only TBM, BM, GRM1 and GRM2 partial mixings are allowed for $U_{R3}$ mixing scheme at 3$\sigma$ CL and the mixing angle $\theta_{34}$ is predicted to be zero for all these cases. There are similar correlations among neutrino oscillation parameters for all viable partial mixings under $U_{R3}$. Fig. 4, shows scatter plots amongst different neutrino mixing angles for TBM partial mixing under $U_{R3}$.
\begin{figure}[h]
\begin{center}
\epsfig{file=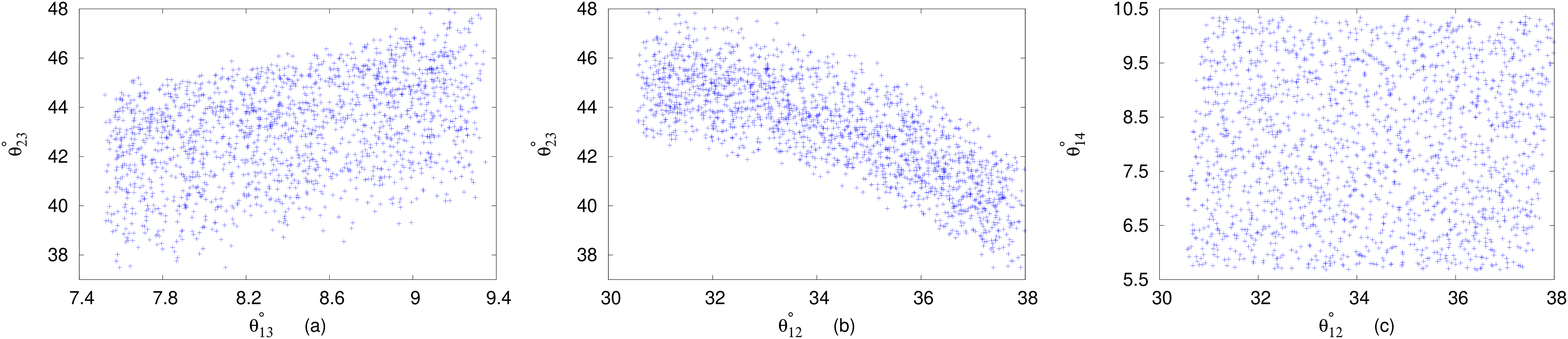, width=16.5cm, height=5.0cm}
\end{center}
\caption{Correlation plots among various neutrino oscillation parameters for $U_{R3}$ mixing scheme with TBM as partial flavour symmetry.}
\end{figure}
\subsection{Mixing Scheme with second row fixed to $N( 1~~b~~a~~0)$}
The general mixing scheme with second row fixed to $N( 1~~b~~a~~0)$ is given by
\begin{small}
\begin{eqnarray}
U_{R2}=\left(
\begin{array}{cccc}
 \frac{c_2 e^{i \phi_2} \left(b c_1 e^{i \phi_1}-a N
   s_1\right)}{\sqrt{b^2+1}} & \frac{c_2 e^{i \phi_2} \left(e^{i
   \phi_1} c_1+a b N s_1\right)}{\sqrt{b^2+1}} & \sqrt{b^2+1}
   c_2 e^{i \phi_2} N s_1 & s_2 \\
 -N & b N & -a N & 0 \\
 \frac{b u-a N x}{\sqrt{b^2+1}} &
   \frac{u+a b N x}{\sqrt{b^2+1}} &
   \sqrt{b^2+1} N x & c_2 s_3 \\
 \frac{b v+a N y}{\sqrt{b^2+1}} &
   \frac{v-a b N y}{\sqrt{b^2+1}} &
   -\sqrt{b^2+1} N y & c_2 c_3 \\
\end{array}
\right),
\end{eqnarray}
\end{small}
where
$u=-c_3 e^{i (\phi_1+\phi_3)} s_1-c_1 e^{i(\phi_1+\phi_2)} s_2 s_3, v=-c_1 c_3 e^{i (\phi_1+\phi_2)} s_2+e^{i
(\phi_1+\phi_3)} s_1 s_3, x=c_1 c_3 e^{i \phi_3}-s_1 s_2 s_3 e^{i \phi_2}, y=c_1 s_3 e^{i \phi_3}+c_3 s_1 s_2 e^{i \phi_2}$.
The three independent conditions $|U_{\mu4}|=0,|U_{\mu3}|=a N$ and $|U_{\mu2}|=b N$ give
\begin{eqnarray}
|\sin\theta_{24}|&=&0,\nonumber\\
\sin^2\theta_{23}&=&\frac{a^2 N^2}{\cos^2\theta_{13}}=\frac{a^2 N^2(1-|U_{e4}|^2)}{1-|U_{e4}|^2-|U_{e3}|^2}, ~ \textrm{and} \\
 \cos\delta_{13}&=&-2 b^2 N^2 \csc(2\theta_{12}) \csc\theta_{13} \csc(2\theta_{23})+\frac{1}{2} \tan\theta_{12} \sin\theta_{13} \tan\theta_{23}+\frac{1}{2} \cot\theta_{12} \csc\theta_{13} \cot\theta_{23},\nonumber
\end{eqnarray}
respectively. This mixing scheme predicts $\theta_{24}=0$ which is not consistent with the recent global (3+1) neutrino oscillation data \cite{li} and is, therefore, phenomenologically ruled out.\\ 
For the partial mixing schemes discussed above, $a$ and $b$ are fixed to the values listed in Table II while the other parameters $\theta_{1}$, $\theta_{2}$, $\theta_{3}$, $\phi_{1}$, $\phi_{2}$ and $\phi_{3}$ are free. We have not considered the parametrization $U_{C3}$ in which the third column will be $N( 0~~a~~b~~0)^T$ and $U_{R1}$ in which the first row will be $N( a~~b~~0~~0)$, since they predict vanishing (1,3) element of the neutrino mixing matrix which is experimentally ruled out.\\
The above parametrizations have six free parameters viz. $\theta_{1}$, $\theta_{2}$, $\theta_{3}$, $\phi_{1}$, $\phi_{2}$ and $\phi_{3}$. Thus, the six neutrino mixing angles $\theta_{13}$, $\theta_{12}$, $\theta_{23}$, $\theta_{14}$, $\theta_{24}, \theta_{34}$ and the three $CP$-violating phases $\delta_{13}, \delta_{14}$ and $\delta_{24}$ can be expressed in terms of six free parameters. The mixing scheme $U_{R2}$ is not viable, since, it leads to a vanishing $|U_{\mu4}|$ contrary to the current neutrino oscillation data given in Table I. Therefore, we have three viable parametrizations viz., $U_{C1}$, $U_{C2}$ and $U_{R3}$. The full allowed parameter space for the mixing schemes $U_{C1}$, $U_{C2}$ and $U_{R3}$ at 3$\sigma$ CL is given in Table III. Table IV gives the allowed ranges of various parameters at 3$\sigma$ CL for the viable partial mixing schemes.       
\begin{table}[h]
\begin{center}
 \begin{tabular}{cccccc}
   \hline
   \hline
   Mixing scheme & $a$ & $b$ &$\theta_{1}$&$\theta_{2}$ &$\theta_{3}$ \\
   \hline
   $U_{C1}$  & $1.4-3.6$ & $0.4-2.2$ &$<14^{\circ}$ & $9^{\circ}-18.5^{\circ}$ & $9^{\circ}-18^{\circ}$\\
   $U_{C2}$  & $0.7-1.5$ & $0.6-1.7$ &$<14.5^{\circ}$ & $6^{\circ}-13^{\circ}$ & $9^{\circ}-12^{\circ}$ \\
   $U_{R3}$  & $1.0-3.6$ & $0.8-2.8$ &$9^{\circ}-15^{\circ}$ & $5.5^{\circ}-10.5^{\circ}$ & $4.0^{\circ}-9.5^{\circ}$\\
    \hline
 \end{tabular}
 \end{center}
 \caption{The experimentally allowed values of various parameters at 3$\sigma$ CL for the mixing schemes $U_{C1},U_{C2}$ and $U_{R3}$. The phases $\phi_1,\phi_2$ and $\phi_3$ can take any value within the range $(0 - 2\pi)$.}
\end{table}
\begin{table}[h]
\begin{small}
\begin{center}
 \begin{tabular}{cccccccc}
   \hline
   \hline
   &  &$\theta_{1}$&$\theta_{2}$ &$\theta_{3}$&$\phi_{1}$&$\phi_{2}$&$\phi_{3}$ \\
   \hline
 $U_{C1}$  & TBM &$<12.5^{\circ}$ & $9.8^{\circ}-18.2^{\circ}$ & $13^{\circ}-17^{\circ}$& $0-360^{\circ}$ & $0-360^{\circ}$ & $50^{\circ}-141^{\circ}$\\
 & & & & & & & $(220^{\circ}-310^{\circ})$\\ 
  & GRM2 &$<12.5^{\circ}$ & $9.6^{\circ}-18^{\circ}$ & $12.9^{\circ}-16.5^{\circ}$& $0-360^{\circ}$ & $0-360^{\circ}$ & $50^{\circ}-150^{\circ}$\\
  & & & & & & & $(215^{\circ}-315^{\circ})$\\    
   \hline
 & TBM &$<14.2^{\circ}$ & $6.8^{\circ}-13^{\circ}$ & $9.2^{\circ}-11.6^{\circ}$& $0-360^{\circ}$ & $0-360^{\circ}$ & $0-360^{\circ}$\\
  & TFH1 &$<14.2^{\circ}$ & $6.8^{\circ}-13^{\circ}$ & $9.2^{\circ}-11.6^{\circ}$& $0-360^{\circ}$ & $0-360^{\circ}$ & $0-360^{\circ}$\\ 
 $U_{C2}$ & TFH2 &$<14.2^{\circ}$ & $6.8^{\circ}-13^{\circ}$ & $9.2^{\circ}-11.6^{\circ}$& $0-360^{\circ}$ & $0-360^{\circ}$ & $0-360^{\circ}$\\
 & HM &$<14.2^{\circ}$ & $6.5^{\circ}-12^{\circ}$ & $8.5^{\circ}-11^{\circ}$& $0-360^{\circ}$ & $0-360^{\circ}$ & $0-360^{\circ}$\\
 & GRM1 &$<14.1^{\circ}$ & $6.6^{\circ}-12.4^{\circ}$ & $8.8^{\circ}-11.1^{\circ}$& $0-360^{\circ}$ & $0-360^{\circ}$ & $0-360^{\circ}$\\
 & GRM2 &$<14^{\circ}$ & $6.8^{\circ}-13^{\circ}$ & $9.3^{\circ}-11.7^{\circ}$& $0-360^{\circ}$ & $0-360^{\circ}$ & $0-360^{\circ}$\\       
   \hline
& TBM &$10.5^{\circ}-13.5^{\circ}$ & $5.5^{\circ}-10.5^{\circ}$ & $4.4^{\circ}-9.5^{\circ}$& $70^{\circ}-130^{\circ}$ & $0-360^{\circ}$ & $0-360^{\circ}$\\
& & & & &$(220^{\circ}-290^{\circ})$& &\\
$U_{R3}$& BM &$10.5^{\circ}-13.5^{\circ}$ & $5.5^{\circ}-10.5^{\circ}$ & $4.4^{\circ}-9.5^{\circ}$& $140^{\circ}-230^{\circ}$ & $0-360^{\circ}$ & $0-360^{\circ}$\\
& GRM1 &$10.5^{\circ}-13.5^{\circ}$ & $5.5^{\circ}-10.5^{\circ}$ & $4.4^{\circ}-9.5^{\circ}$& $40^{\circ}-105^{\circ}$ & $0-360^{\circ}$ & $0-360^{\circ}$\\
& & & & &$(255^{\circ}-325^{\circ})$& &\\
& GRM2 &$11.4^{\circ}-14.5^{\circ}$ & $5.5^{\circ}-10.5^{\circ}$ & $4.4^{\circ}-9.5^{\circ}$& $80^{\circ}-135^{\circ}$ & $0-360^{\circ}$ & $0-360^{\circ}$\\
& & & & &$(225^{\circ}-280^{\circ})$& &\\
\hline 
 \end{tabular}
 \end{center}
 \caption{The experimentally allowed values of various parameters at 3$\sigma$ CL for different partial mixing schemes.}
\end{small}
\end{table}  

\section{General $4\times4$ Mixing Schemes with one row or one column fixed}
In this section, we discuss $4\times4$ partial mixing schemes with one column or one row fixed and none of the mixing matrix element equal to zero. Here, we study the phenomenology of $4\times4$ mixing scheme keeping the first column or first row fixed.    

\subsection{Mixing Scheme with one column fixed to $N( a~~b~~c~~1)^T$}
Here, we discuss the possibility of having any one of the columns of $4\times4$ mixing matrix fixed to $N( a~~b~~c~~1)^T$. Any column of the mixing matrix fixed to $N( a~~b~~c~~1)^T$ gives three independent conditions on the magnitudes of the elements of mixing matrix viz.,
\begin{equation}
|U_{e1}|=a N,~~~~|U_{\mu 1}|=b N~~~ \textrm{and}~~~|U_{c N}|=c N
\end{equation}
where $N=1/\sqrt{1+a^2+b^2+c^2}$ is the normalization factor. Here, we consider the general mixing matrix of the form
\begin{equation}
U^{\prime} = P^{\prime}R_{14}(\theta_{4},\phi_{4})R_{12}(\theta_{1},\phi_{1})R_{13}(\theta_{3},\phi_{3})R_{24}(\theta_{5},\phi_{5})R_{23}(\theta_{2},\phi_{2})R_{34}(\theta_{6},\phi_{6}) P
\end{equation}
where $R_{ij}(\theta_{k},\phi_{l})$ is the rotation matrix in the $i$-$j$ plane with $\phi_{l}$ as the phase angle. $P^{\prime}$ and $P$ are two diagonal phase matrices. In the above mixing matrix, the three phases ($\phi_1,\phi_3,\phi_4$) can be associated with the Majorana-type CP-violating phases and can be extracted out. Using three conditions from Eq. (41), the general mixing matrix with first column fixed to $N( a~~b~~c~~1)^T$ becomes  
\begin{small}
\begin{equation}
U^{\prime}_{C1}=\left(
\begin{array}{cccc}
 a N & U_{e2} & U_{e3} & U_{e4} \\
 -b N & U_{\mu2} & U_{\mu3} & U_{\mu4} \\
 -c N & U_{\tau2} & U_{\tau3} & U_{\tau4} \\
 -N & U_{s2} & U_{s3} & U_{s4} \\
\end{array}
\right)
\end{equation}
where
\begin{eqnarray}
U_{e2}&=&\cos\theta_{2} \left(\sin\theta_{1} \cos\theta_{4} \cos\theta_{5}-e^{i \phi_{5}} \sin\theta_{4} \sin\theta_{5}\right)-e^{i \phi_{2}} \cos\theta_{1} \sin\theta_{2}\sin\theta_{3} \cos\theta_{4},\nonumber\\
U_{e3}&=&\cos\theta_{6} \left(\cos\theta_{1} \cos\theta_{2} \sin\theta_{3} \cos\theta_{4}+e^{-i \phi_{2}} \sin\theta_{2} \left(\sin\theta_{1} \cos\theta_{4} \cos\theta_{5}-e^{i\phi_{5}} \sin\theta_{4} \sin\theta_{5}\right)\right)-\nonumber\\
&& e^{i\phi_{6}} \sin\theta_{6} \left(\sin\theta_{4} \cos\theta_{5}+e^{-i\phi_{5}} \sin\theta_{1} \cos\theta_{4} \sin\theta_{5}\right),\nonumber\\
U_{e4}&=&e^{-i \phi_{6}} \sin \theta_{6} \left(\cos\theta_{1} \cos\theta_{2} \sin \theta_{3} \cos\theta_{4}+e^{-i
\phi_{2}} \sin\theta_{2} \left(\sin\theta_{1} \cos\theta_{4} \cos\theta_{5}-e^{i\phi_{5}} \sin\theta_{4} \sin\theta_{5}\right)\right)+\nonumber\\
&& \cos\theta_{6} \left(\sin\theta_{4} \cos\theta_{5}+e^{-i \phi_{5}} \sin\theta_{1} \cos\theta_{4} \sin\theta_{5}\right),\nonumber\\
U_{\mu2}&=&\cos \theta_{1} \cos \theta_{2} \cos\theta_{5}+e^{i\phi_{2}} \sin\theta_{1} \sin\theta_{2} \sin\theta_{3},\nonumber\\
U_{\mu3}&=&\cos\theta_{6} \left(-\sin\theta_{1} \cos\theta_{2} \sin\theta_{3}+e^{-i\phi_{2}} \cos\theta_{1} \sin\theta_{2} \cos\theta_{5}\right)-\cos\theta_{1} \sin\theta_{5} \sin\theta_{6} e^{i\phi_{6}-i\phi_{5}},\nonumber\\
U_{\mu4}&=&e^{-i\phi_{6}} \sin\theta_{6} \left(-\sin\theta_{1} \cos\theta_{2} \sin\theta_{3}+e^{-i\phi_{2}} \cos\theta_{1} \sin\theta_{2} \cos\theta_{5}\right)+e^{-i
\phi_{5}} \cos\theta_{1} \sin\theta_{5} \cos\theta_{6},\nonumber\\
U_{\tau2}&=&-e^{i\phi_{2}} \sin\theta_{2} \cos\theta_{3},\nonumber\\
U_{\tau3}&=&\cos\theta_{2} \cos\theta_{3} \cos\theta_{6},\nonumber\\
U_{\tau4}&=&e^{-i\phi_{6}} \cos\theta_{2} \cos\theta_{3} \sin\theta_{6},\nonumber\\
U_{s2}&=&e^{i\phi_{2}} \cos\theta_{1} \sin\theta_{2} \sin\theta_{3} \sin\theta_{4}+\cos\theta_{2} \left(\sin\theta_{1} \sin\theta_{4} (-\cos\theta_{5})-e^{i\phi_{5}} \cos\theta_{4} \sin\theta_{5}\right),\nonumber\\
U_{s3}&=&\cos\theta_{6}\left(-\cos\theta_{1} \cos\theta_{2} \sin\theta_{3} \sin\theta_{4}+e^{-i \phi_{2}} \sin\theta_{2} \left(\sin\theta_{1} \sin\theta_{4} (-\cos\theta_{5})-e^{i\phi_{5}} \cos\theta_{4} \sin\theta_{5}\right)\right)-\nonumber\\
&& e^{i\phi_{6}} \sin\theta_{6}
   \left(\cos\theta_{4} \cos\theta_{5}-e^{-i \phi_{5}} \sin\theta_{1} \sin\theta_{4})\sin\theta_{5}\right),\nonumber\\
U_{s4}&=&e^{-i\phi_{6}} \sin\theta_{6} \left(-\cos\theta_{1} \cos\theta_{2} \sin\theta_{3} \sin\theta_{4}+e^{-i
\phi_{2}} \sin\theta_{2} \left(\sin\theta_{1} \sin\theta_{4} (-\cos\theta_{5})-e^{i\phi_{5}} \cos\theta_{4} \sin\theta_{5}\right)\right)+\nonumber\\
&& \cos\theta_{6}\left(\cos\theta_{4} \cos\theta_{5}-e^{-i\phi_{5}} \sin\theta_{1} \sin\theta_{4} \sin\theta_{5}\right)
\end{eqnarray}
\end{small}
with
\begin{eqnarray}
\sin\theta_{1}&&=\frac{b N}{\sqrt{1-c^2 N^2}},\nonumber\\
\sin\theta_{3}&&=c N,\nonumber\\
\cos\theta_{4}&&=\frac{a N}{\sqrt{1-b^2 N^2-c^2 N^2}}.
\end{eqnarray}

\subsection{Mixing Scheme with one row fixed to $N( a~~b~~c~~1)$}
A mixing scheme with first row fixed to $N( a~~b~~c~~1)$ leads to the following three independent condition on the magnitudes of the elements of neutrino mixing matrix:
\begin{eqnarray}
|U_{e1}|=a N,~~~|U_{e2}|=bN,~~~|U_{e3}|=cN
\end{eqnarray}
where $N=1/\sqrt{1+a^2+b^2+c^2}$ is the normalization factor. Considering the mixing scheme of the form
\begin{equation}
U^{\prime\prime}=R_{34}(\theta_{6},\phi_{6})R_{24}(\theta_{5},\phi_{5})R_{23}(\theta_{2},\phi_{2})R_{14}(\theta_{4})R_{12}(\theta_{1})R_{13}(\theta_{3})
\end{equation}
where the three phases ($\phi_1,\phi_3,\phi_4$) associated with the Majorana-type CP-violating phases can be extracted out. Using conditions from Eq. (46), a neutrino mixing scheme with first row fixed is given by   
\begin{small}
\begin{equation}
U^{\prime\prime}_{R1}=\left(
\begin{array}{cccc}
 a N & b N & c N & N \\
 U_{\mu1} & U_{\mu2} & U_{\mu3} & U_{\mu4} \\
 U_{\tau1} & U_{\tau2} & U_{\tau3} & U_{\tau4} \\
 U_{s1} & U_{s2} & U_{s3} & U_{s4} \\
\end{array}
\right)
\end{equation}
where
\begin{eqnarray}
U_{\mu1}&=&\cos\theta_{3} \left(\sin\theta_{1} (-\cos\theta_{2}) \cos
   \theta_{5}-e^{-i\phi_{5}} \cos\theta_{1} \sin\theta_{4} \sin\theta_{5}\right)-e^{-i \phi_{2}} \sin
   \theta_{2} \sin\theta_{3} \cos\theta_{5},\nonumber\\
U_{\mu2}&=&\cos\theta_{1} \cos\theta_{2} \cos\theta_{5}-e^{-i
   \phi_{5}} \sin\theta_{1} \sin\theta_{4} \sin\theta_{5},\nonumber\\
U_{\mu3}&=&\sin\theta_{3} \left(\sin\theta_{1} (-\cos\theta_{2}) \cos
   (\theta_{5}-e^{-i \phi_{5}} \cos\theta_{1} \sin\theta_{4} \sin\theta_{5}\right)+e^{-i \phi_{2}} \sin
   \theta_{2} \cos\theta_{3} \cos\theta_{5},\nonumber\\
U_{\mu4}&=&e^{-i\phi_{5}} \cos\theta_{4} \sin\theta_{5},\nonumber\\
U_{\tau1}&=&\cos\theta_{3} (-\sin\theta_{1} \left(-\cos\theta_{2}
   \sin\theta_{5} \sin\theta_{6} e^{i(\phi_{5}-\phi_{6})}-e^{i \phi_{2}} \sin\theta_{2} \cos\theta_{6}\right)-\nonumber\\
   && e^{-i \phi_{6}} \cos\theta_{1} \sin\theta_{4} \cos\theta_{5} \sin\theta_{6})-\sin\theta_{3}
   \left(\cos\theta_{2} \cos\theta_{6}-\sin\theta_{2} \sin\theta_{5} \sin\theta_{6} e^{i(- \phi_{2}+ \phi_{5}-\phi_{6})}\right),\nonumber\\
U_{\tau2}&=&\cos\theta_{1} \left(-\cos\theta_{2} \sin\theta_{5} \sin\theta_{6} e^{i \phi_{5}-i\phi_{6}}-e^{i \phi_{2}} \sin\theta_{2} \cos\theta_{6}\right)-e^{-i \phi_{6}} \sin\theta_{1} \sin\theta_{4} \cos\theta_{5} \sin\theta_{6},\nonumber\\
U_{\tau3}&=&\sin\theta_{3} (-\sin\theta_{1} \left(-\cos\theta_{2}
 \sin\theta_{5} \sin\theta_{6} e^{i\phi_{5}-i \phi_{6}}-e^{i \phi_{2}} \sin \theta_{2} \cos\theta_{6}\right)-\nonumber\\
 && e^{-i \phi_{6}} \cos\theta_{1} \sin\theta_{4} \cos\theta_{5} \sin\theta_{6})+\cos\theta_{3} \left(\cos\theta_{2} \cos \theta_{6}-\sin\theta_{2} \sin\theta_{5} \sin\theta_{6} e^{i(-\phi_{2}+\phi_{5}-\phi_{6})}\right),\nonumber\\
U_{\tau4}&=&e^{-i\phi_{6}} \cos\theta_{4} \cos\theta_{5} \sin\theta_{6},\nonumber\\
U_{s1}&=&\cos\theta_{3} \left(-\cos\theta_{1} \sin\theta_{4} \cos\theta_{5} \cos\theta_{6}-\sin\theta_{1} \left(\sin\theta_{2} \sin\theta_{6} e^{i(\phi_{2}+\phi_{6})}-e^{i\phi_{5}} \cos\theta_{2} \sin\theta_{5} \cos\theta_{6}\right)\right)-\nonumber\\
&& \sin\theta_{3} \left(\sin\theta_{2} \sin\theta_{5} \cos\theta_{6} \left(-e^{i(\phi_{5}-\phi_{2})}\right)-e^{i\phi_{6}} \cos\theta_{2} \sin\theta_{6}\right),\nonumber\\
U_{s2}&=&-\sin\theta_{1} \sin\theta_{4} \cos\theta_{5} \cos\theta_{6}+\cos\theta_{1} \left(\sin\theta_{2} \sin\theta_{6} e^{i(\phi_{2}+ \phi_{6})}-e^{i \phi_{5}} \cos\theta_{2} \sin\theta_{5} \cos\theta_{6}\right),\nonumber\\
U_{s3}&=&\sin\theta_{3} \left(-\cos\theta_{1} \sin\theta_{4} \cos\theta_{5} \cos\theta_{6}-\sin\theta_{1} \left(\sin
\theta_{2} \sin\theta_{6} e^{i(\phi_{2}+ \phi_{6})}-e^{i \phi_{5}} \cos\theta_{2} \sin\theta_{5} \cos\theta_{6}\right)\right)+\nonumber\\
&& \cos\theta_{3} \left(\sin\theta_{2} \sin\theta_{5} \cos\theta_{6} \left(-e^{i(\phi_{5}-\phi_{2})}\right)-e^{i \phi_{6}} \cos\theta_{2} \sin\theta_{6}\right),\nonumber\\
U_{s4}&=&\cos\theta_{4} \cos\theta_{5} \cos\theta_{6}
\end{eqnarray}
with
\begin{eqnarray}
\sin\theta_{1}&=&\frac{b}{\sqrt{a^2+b^2+c^2}},\nonumber\\
\sin\theta_{3}&=&\frac{c}{\sqrt{a^2+c^2}},\nonumber\\
\cos\theta_{4}&=&N \sqrt{a^2+b^2+c^2}~.
\end{eqnarray}
\end{small}

For the numerical analysis, about $10^7\sim10^8$ points are generated randomly. Comparing Eq. (6) with Eq. (43) for $U^{\prime}_{C1}$ mixing and Eq. (48) for $U^{\prime\prime}_{R1}$ mixing, the corresponding six neutrino mixing angles can be calculated by using Eq. (7). The parameters $\theta_{2},\theta_{5},\theta_{6}$ are generated randomly within the range $(0 - \frac{\pi}{2}$) and phases $\phi_{3},\phi_{5},\phi_{6}$ are generated within the range $(0 - 2\pi$). For $U^{\prime\prime}_{R1}$ mixing scheme the parameters $a$, $b$ and $c$ are varied randomly and for $U^{\prime}_{C1}$ mixing these parameters are varied within the ranges: $a<100,b<100,c<100$. Using available experimental constraints on neutrino oscillation parameters, we obtain the correlations among various parameters as shown in Fig. 5 for $U^{\prime}_{C1}$ mixing and Fig. 6 for $U^{\prime\prime}_{R1}$ mixing. The allowed ranges for various parameters at 3$\sigma$ CL are given in Table V. By following the same procedure, we can obtain the $4\times4$ mixing matrix with any column (row) fixed to $N( a~~b~~c~~1)^T$ ($N( a~~b~~c~~1)$). 
\begin{figure}[h]
\begin{center}
\epsfig{file=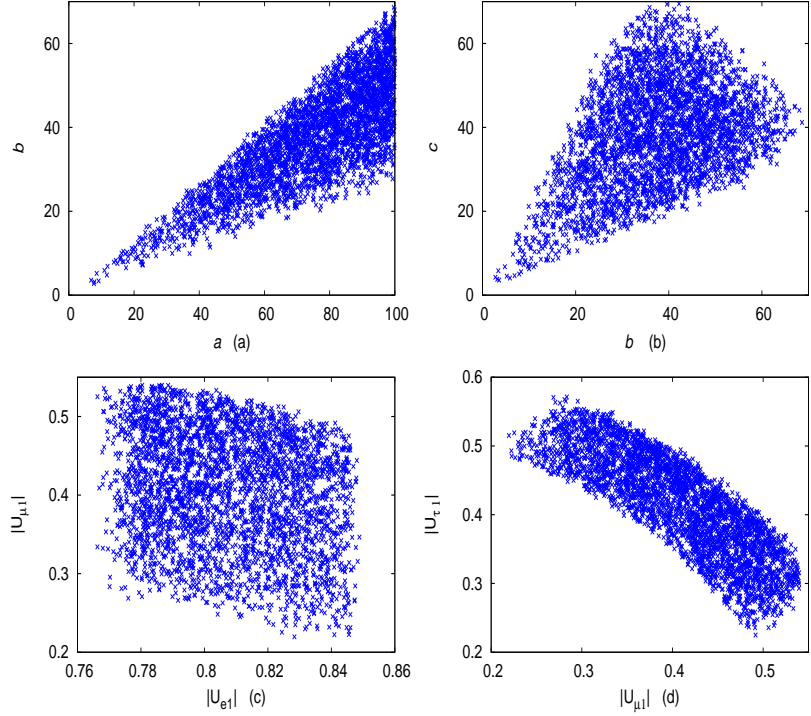, width=11.cm, height=10.0cm}
\end{center}
\caption{Correlation plots among parameters $a,b,c$  and among neutrino mixing matrix elements for $U^{\prime}_{C1}$ mixing.}
\end{figure}

\begin{figure}[h!]
\begin{center}
\epsfig{file=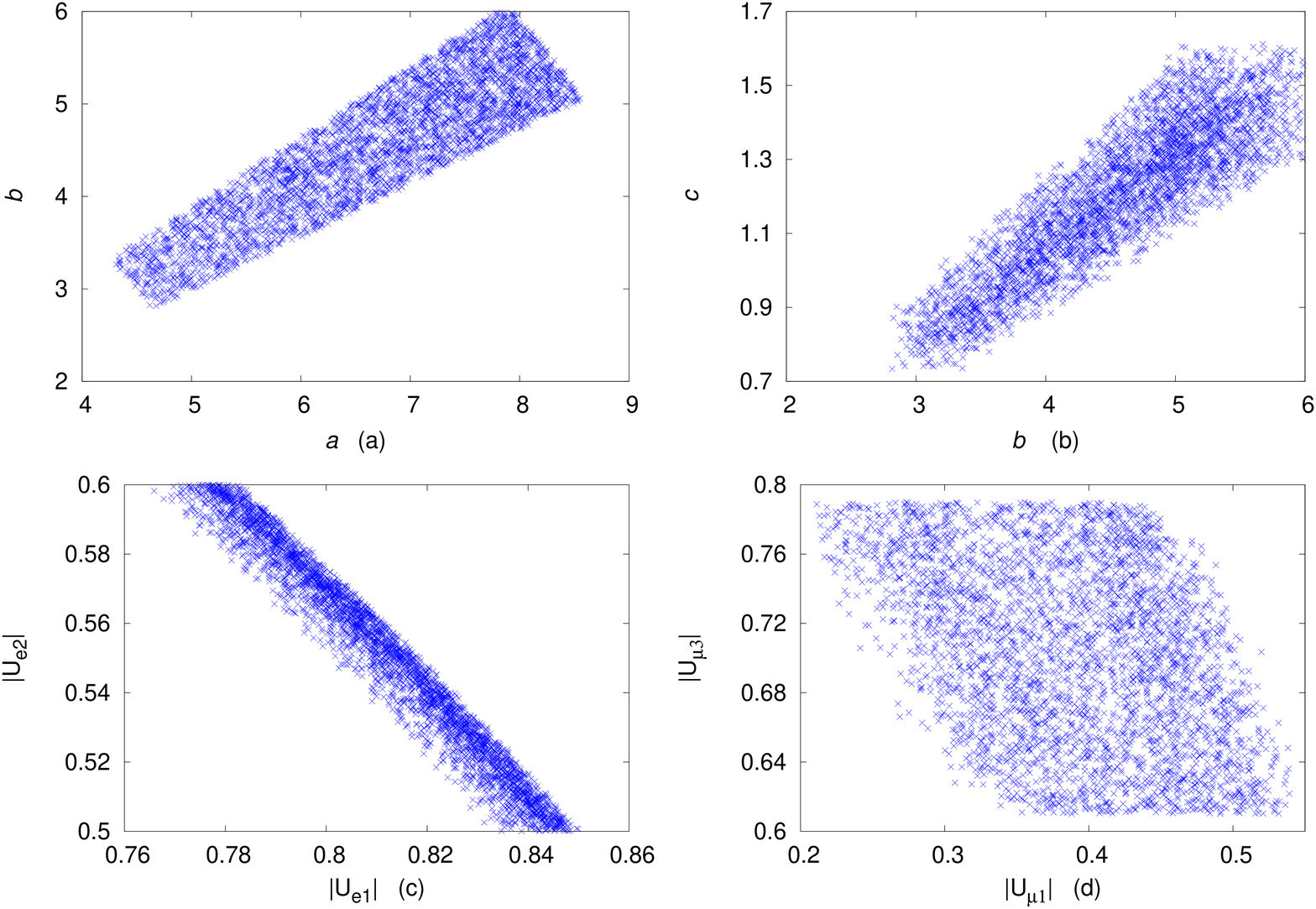, width=11.0cm, height=10.0cm}
\end{center}
\caption{Correlation plots among parameters $a,b,c$ and among neutrino mixing matrix elements for $U^{\prime\prime}_{R1}$ mixing.}
\end{figure}

\begin{table}[h]
\begin{center}
 \begin{tabular}{cccccccc}
   \hline
   \hline
   Mixing scheme & $a$ & $b$ & $c$ & $\theta_{2}$&$\theta_{5}$ &$\theta_{6}$ & $\phi_{2}$ \\
   \hline
   $U^{\prime}_{C1}$  & - &-&-&$25^{\circ}-55^{\circ}$ & $<30^{\circ}$ & $<20^{\circ}$& $140^{\circ}-220^{\circ}$\\
   $U^{\prime\prime}_{R1}$  & $4.0-9.0$ & $2.5-6.5$ & $0.7-1.7$ &$30^{\circ}-60^{\circ}$ & $4^{\circ}-9.5^{\circ}$ & $<12^{\circ}$& $0^{\circ}-360^{\circ}$ \\
    \hline
 \end{tabular}
 \end{center}
 \caption{The experimentally allowed values of various parameters at 3$\sigma$ CL for $U^{\prime}_{C1}$ and $U^{\prime\prime}_{R1}$ mixing schemes. The phases $\phi_5$ and $\phi_6$ can take any value within the range $(0 - 2\pi)$.}
\end{table}

\section{Summary}
In this work, we presented new mixing schemes for (3+1) neutrinos. These include partial mixing schemes having one column or one row of the $4\times4$ mixing matrix fixed to that of the popular mixing schemes such as TBM, BM, DM, HM, GRM1, GRM2, TFH1 and TFH2. These mixing schemes are useful to describe active-active and active-sterile neutrino mixings. These partial mixing schemes are obtained by modifications to exact mixing schemes such as TBM, DM, BM, etc. and can be factored into two parts: $V(a,b)$ and $R(\theta,\phi)$. The $V(a,b)$ part represents one of the popular mixing scheme like TBM etc. having flavor symmetric origin and the $R(\theta,\phi)$ part can be considered as a correction to $V(a,b)$. We calculated the experimentally allowed parameter space for the parameters $a$ and $b$. For $U_{C1}$ mixing with first column fixed, only TBM, GRM2 partial mixing schemes are allowed at 3$\sigma$ CL while for $U_{C2}$ mixing with second column fixed, partial mixings TBM, HM, TFH1, TFH2, GRM1 and GRM2 are allowed at 3$\sigma$ CL. For $U_{R3}$ mixing having third row fixed, the allowed partial mixings are: TBM, BM, GRM1, and GRM2. We, also, studied the phenomenology of general mixing schemes for $(3 + 1)$ neutrinos with one column or one row fixed such that the fixed column or row has no vanishing element. 

\textbf{\textit{\Large{Acknowledgements}}}\\
The research work of S. D. is supported by the Council of Scientific and Industrial Research (CSIR), Government of India, New Delhi vide grant No. 03(1333)/15/EMR-II. S. D. gratefully acknowledges the kind hospitality provided by IUCAA, Pune. R. R. G. acknowledges the financial support provided by the Council of Scientific and Industrial Research (CSIR), Government of India, New Delhi vide grant No. 13(8949-A)/2017-Pool.

\section{Appendix:  The parametrization $U_{C1}$}
Here, we derive the $4\times4$ mixing matrix with first column fixed to 
$( aN, bN, N, 0)^T$:
\begin{equation}
U_{C1}=\left(
\begin{array}{cccc}
 a N & u_1 & v_1 & w_1 \\
 -b N & u_2 & v_2 & w_2 \\
 -N & u_3 & v_3 & w_3 \\
 0 & u_4 & v_4 & w_4
\end{array}
\right).
\end{equation}
The unknown mixing matrix elements can be written as sum of real and imaginary terms and using the orthogonality of columns, we obtain
\begin{small}
\begin{equation}
U_{C1}=\left(
\begin{array}{cccc}
 a N & x_1+i y_1 & x_2+i y_2 & x_5+i y_5 \\
 -b N & x_3+i y_3 & x_4+i y_4 & x_6+i y_6 \\
 -N & a x_1-b x_3+i (a y_1-b y_3) & a x_2-b x_4+i (a
   y_2-b y_4) & a x_5-b x_6+i (a y_5-b y_6) \\
 0 & x_7+i y_7 & x_8+i y_8 & x_9+i y_9
\end{array}
\right).
\end{equation}
\end{small}
The parameters $y_{1}$, $y_{2}$ and $y_{5}$ can be related to Majorana phases which can be factored out as a Majorana phase matrix. Therefore, substituting $y_{1}=y_{2}=y_{5}=0$ in above equation and using the unitarity constraints $U U^{\dagger}=U^{\dagger} U=1$, we get
\begin{small}
\begin{eqnarray}
x_{5}&=&\sqrt{-a^2 N^2-x_1^2-x_2^2+1},\nonumber\\
x_{4}&=&\frac{a b N^2 x_2-\frac{\sqrt{d}}{2}-x_1 x_2
   x_3}{x_2^2+x_5^2},\nonumber\\
x_{6}&=&\frac{a b N^2-x_1 x_3-x_2 x_4}{x_5},\nonumber\\
y_{6}&=&-\frac{x_1 y_3+x_2 y_4}{x_5},\nonumber\\
y_{7}&=&\sqrt{-\frac{f}{g}},\\
x_{8}&=&\frac{x_1 \left(-x_2 x_7 \left(x_6^2+y_6^2\right)+x_4
   x_5 (x_6 x_7+y_6 y_7)+x_5 y_4 (x_7
   y_6-x_6 y_7)\right)}{e}+\nonumber\\
   &&\frac{x_3 x_5 (x_2 x_6
   x_7-x_2 y_6 y_7-x_4 x_5 x_7+x_5
   y_4 y_7)+x_5 y_3 (x_2 (x_6 y_7+x_7
   y_6)-x_5 (x_4 y_7+x_7 y_4))}{e},\nonumber\\
y_{8}&=&\frac{x_2 \left(x_5 (x_3 x_6 y_7+x_3 x_7
   y_6-x_6 x_7 y_3+y_3 y_6 y_7)-x_1
   y_7 \left(x_6^2+y_6^2\right)\right)}{e}+\nonumber\\
   &&\frac{x_4 x_5 (x_1
   x_6 y_7-x_1 x_7 y_6-x_3 x_5
   y_7+x_5 x_7 y_3)+x_5 y_4 (x_1 x_6
   x_7+x_1 y_6 y_7-x_3 x_5 x_7-x_5
   y_3 y_7)}{e},\nonumber\\
x_{9}&=&-\frac{x_3 x_6 x_7-x_3 y_6 y_7+x_4
   x_6 x_8-x_4 y_6 y_8+x_6 y_3
   y_7+x_6 y_4 y_8+x_7 y_3 y_6+x_8
   y_4 y_6}{x_6^2+y_6^2},\nonumber\\
y_{9}&=&\frac{x_6 (-x_3 y_7-x_4 y_8+x_7 y_3+x_8
   y_4)-y_6 (x_3 x_7+x_4 x_8+y_3
   y_7+y_4 y_8)}{x_6^2+y_6^2},\nonumber   
\end{eqnarray}
\end{small} 
where
\begin{small}
\begin{eqnarray}
d&=&\left(2 a b N^2 x_2-2 x_1 x_2 x_3\right)^2-4
   \left(x_2^2+x_5^2\right)\nonumber \\
   && \left(a^2 \left(b^2 N^4-N^2 x_5^2\right)-2
   a b N^2 x_1 x_3-N^2 x_5^2+x_1^2
   \left(x_3^2+y_3^2\right)+
   2 x_1 x_2 y_3
   y_4+x_2^2 y_4^2+x_3^2 x_5^2+x_5^2
   y_3^2+x_5^2 y_4^2\right),\nonumber \\
e&=&x_2^2 \left(x_6^2+y_6^2\right)-2 x_2 x_4 x_5
   x_6-2 x_2 x_5 y_4 y_6+x_4^2
   x_5^2+x_5^2 y_4^2,\\
f&=&x_4^2 \left(x_1^2 x_7^2+x_5^2
   \left(x_7^2-1\right)\right)+x_1^2 x_6^2 x_7^2+x_1^2
   x_7^2 y_4^2+x_1^2 x_7^2 y_6^2-2 x_2 x_4
   \left(x_1 x_3 x_7^2+x_5 x_6
   \left(x_7^2-1\right)\right)- \nonumber\\
   && 2 x_2 y_4 \left(x_1 x_7^2
   y_3+x_5 \left(x_7^2-1\right) y_6\right)-2 x_1 x_3
   x_5 x_6 x_7^2-2 x_1 x_5 x_7^2 y_3
   y_6+ \nonumber\\
   && x_2^2 \left(x_3^2 x_7^2+x_6^2
   \left(x_7^2-1\right)+
    x_7^2 y_3^2+x_7^2
   y_6^2-y_6^2\right)+x_3^2 x_5^2 x_7^2+x_5^2
   x_7^2 y_3^2+x_5^2 x_7^2 y_4^2-x_5^2 y_4^2, \nonumber\\
g&=&x_1^2 \left(x_4^2+x_6^2+y_4^2+y_6^2\right)-2 x_2
   (x_1 x_3 x_4+x_1 y_3 y_4+x_4 x_5
   x_6+x_5 y_4 y_6)- \nonumber\\
   && 2 x_1 x_5 (x_3
   x_6+y_3 y_6)+x_2^2
   \left(x_3^2+x_6^2+y_3^2+y_6^2\right)+x_5^2
   \left(x_3^2+x_4^2+y_3^2+y_4^2\right). \nonumber
\end{eqnarray}
\end{small}
In total, there are six free parameters in the mixing matrix $U_{C1}$, viz. $x_{1}, x_{2}, x_{3}, y_{3}, y_{4}$ and $x_{7}$. These parameters can be further reparametrized in terms of six angles $\theta_{1}, \theta_{2}, \theta_{3}, \phi_{1}, \phi_{2}$ and $\phi_{3}$ as
\begin{eqnarray}
x_{1}&=&\sqrt{b^2+1} N \cos\theta_2 \cos\theta_3,\nonumber \\
x_{2}&=&\sqrt{b^2+1} N \cos\theta_2 \sin\theta_3,\nonumber \\
x_{3}&=&\frac{a b N \cos\theta_2 \cos\theta_3+\sin\theta_1
   \sin\theta_2 \cos\theta_3 \cos (\phi_1-\phi_2)+\cos\theta_1 \sin\theta_3 \cos\phi_3}{\sqrt{b^2+1}},\nonumber \\
y_{3}&=&\frac{\cos\theta_1 \sin\theta_3 \sin\phi_3-\sin\theta_1 \sin\theta_2 \cos\theta_3 \sin(\phi_1-\phi_2)}{\sqrt{b^2+1}},\\
y_{4}&=&-\frac{\sin\theta_1 \sin\theta_2 \sin\theta_3 \sin(\phi_1-\phi_2)+\cos\theta_1 \cos\theta_3 
   \sin\phi_3}{\sqrt{b^2+1}},\nonumber\\
x_{7}&=&\sin\theta_1 \sin\theta_3 \cos(\phi_1+\phi_3)-\cos\theta_1 \sin\theta_2 \cos\theta_3 \cos \phi_2. \nonumber
\end{eqnarray} 
With these redefinitions, the most general mixing matrix of type $U_{C1}$ becomes
\begin{scriptsize}
\begin{eqnarray}
U_{C1}=\left(
\begin{array}{cccc}
 a N & \sqrt{b^2+1} c_2 c_3 N & \sqrt{b^2+1} c_2 N s_3 &
   \sqrt{b^2+1} N s_2 \\
 -b N & \frac{\left(a b c_2 c_3 N+c_3 e^{i (\phi_2- \phi_1)} s_1 s_2+c_1 e^{i \phi_3} s_3\right)}{\sqrt{b^2+1}} & \frac{
   \left(-c_1 e^{i \phi_3} c_3+a b c_2 N s_3+e^{i (\phi_2-\phi_1)} s_1 s_2 s_3 \right)}{\sqrt{b^2+1}} & \frac{\left(a b N s_2-c_2 e^{i (\phi_2-\phi_1)}  s_1\right)}{\sqrt{b^2+1}} \\
 -N & \frac{\left(a c_2 c_3 N-b \left(c_3 e^{i (\phi_2-\phi_1)} s_1 s_2+c_1 e^{i \phi_3} s_3\right)\right)}{\sqrt{b^2+1}} & \frac{\left(b c_1 e^{i \phi_3} c_3+a c_2 N s_3-b e^{i (\phi_2-\phi_1)} s_1 s_2
   s_3\right)}{\sqrt{b^2+1}} & \frac{\left(b c_2 e^{i
   (\phi_2-\phi_1)} s_1+a N s_2\right)}{\sqrt{b^2+1}}
   \\
 0 & e^{i (\phi_1+\phi_3)} s_1 s_3-c_1 c_3 e^{i
   \phi_2} s_2 & -c_3 e^{i (\phi_1+\phi_3)}
   s_1-c_1 e^{i \phi_2} s_2 s_3 & c_1 c_2
   e^{i \phi_2} \\
\end{array}
\right)P.
\end{eqnarray}
\end{scriptsize}

\end{document}